\newcommand{\mum}{\,$\rm \,\rm\mu m$\xspace}
\newcommand*{\av}{$A_{V}$\xspace}
\newcommand*{\arcsecf}{\hbox{$.\!\!^{\prime\prime}$}}

\documentclass{aa}  

\usepackage{graphicx}
\graphicspath{ {./figures/} }
\usepackage{txfonts}
\usepackage{xcolor}

\newcommand{\orcid}[1]{}
\usepackage[]{hyperref}

\begin{document} 

\title{HELM's deep: Highly Extincted Low-Mass galaxies seen by JWST}

\author{L.~Bisigello\orcid{0000-0003-0492-4924}\thanks{\email{laura.bisigello@inaf.it}}\inst{\ref{aff1}}
\and G.~Gandolfi\orcid{0000-0002-9415-2296}\inst{\ref{aff2},\ref{aff1}}
\and A. Grazian\inst{\ref{aff1}}
\and G.~Rodighiero\orcid{0000-0002-9415-2296}\inst{\ref{aff2},\ref{aff1}}
\and G.~Girardi\inst{\ref{aff1},\ref{aff2}}
\and A.~Renzini\inst{\ref{aff1}}
\and A.~Vietri\inst{\ref{aff2}}
\and E.~McGrath \inst{\ref{aff6}}
\and B.~Holwerda\inst{\ref{aff9}}
\and Abdurro'uf\inst{\ref{aff4},\ref{aff5},\ref{aff22}}
\and M.~Castellano\inst{\ref{aff7}}
\and M.~Giulietti\inst{\ref{aff3}}
\and C.~Gruppioni\inst{\ref{aff8}}
\and N.~Hathi\inst{\ref{aff5}}
\and A.~M.~Koekemoer\inst{\ref{aff5}}
\and R.~Lucas\inst{\ref{aff5}}
\and F.~Pacucci\inst{\ref{aff10},\ref{aff11}}
\and P.~G.~P\'erez-Gonz\'alez\inst{\ref{aff12}}
\and L.~Y.~A.~Yung\orcid{0000-0003-3466-035X}\inst{\ref{aff5}}
\and P. Arrabal Haro\inst{\ref{aff13}}
\and B. E. Backhaus\inst{\ref{aff21}}
\and M. Bagley\inst{\ref{aff13},\ref{aff14}}
\and M. Dickinson\inst{\ref{aff15}}
\and S. Finkelstein\inst{\ref{aff14},\ref{aff16}}
\and J. Kartaltepe\inst{\ref{aff17}}
\and A. Kirkpatrick\inst{\ref{aff21}}
\and C. Papovich\inst{\ref{aff18},\ref{aff19}}
\and N. Pirzkal\inst{\ref{aff20}}
}

\institute{INAF-Osservatorio Astronomico di Padova, Via dell'Osservatorio 5, 35122 Padova, Italy\label{aff1}
\and 
Dipartimento di Fisica e Astronomia "G. Galilei", Universit\`a di Padova, Via Marzolo 8, 35131 Padova, Italy\label{aff2}
\and 
INAF, Istituto di Radioastronomia, Via Piero Gobetti 101, 40129 Bologna, Italy\label{aff3}
\and
Department of Physics and Astronomy, Colby College, Waterville, ME 04901, USA\label{aff6}
\and
Department of Physics and Astronomy, The Johns Hopkins University, 3400 N Charles St., Baltimore, MD 21218, USA\label{aff4}
\and
INAF - Osservatorio Astronomico di Roma, via di Frascati 33, 00078 Monte Porzio Catone, Italy\label{aff7}
\and
Space Telescope Science Institute (STScI), 3700 San Martin Drive, Baltimore, MD 21218, USA\label{aff5}
\and
INAF-Osservatorio di Astrofisica e Scienza dello Spazio, Via Gobetti 93/3, 40129 Bologna, Italy\label{aff8}
\and
Physics \& Astronomy Department, University of Louisville, 40292 KY, Louisville, USA\label{aff9}
\and
Center for Astrophysics $\vert$ Harvard \& Smithsonian, 60 Garden St, Cambridge, MA 02138, USA\label{aff10}
\and
Black Hole Initiative, Harvard University, 20 Garden St, Cambridge, MA 02138, USA\label{aff11}
\and
Centro de Astrobiolog\'{\i}a (CAB), CSIC-INTA, Ctra. de Ajalvir km 4, Torrej\'on de Ardoz, E-28850, Madrid, Spain\label{aff12}
\and
Astrophysics Science Division, NASA Goddard Space Flight Center, 8800 Greenbelt Rd, Greenbelt, MD 20771, USA\label{aff13}
\and
Department of Astronomy, The University of Texas at Austin, Austin, TX, USA\label{aff14}
\and
NSF's National Optical-Infrared Astronomy Research Laboratory, 950 N. Cherry Ave., Tucson, AZ 85719, USA\label{aff15}
\and
Cosmic Frontier Center, The University of Texas at Austin, Austin, TX, USA\label{aff16}
\and
Laboratory for Multiwavelength Astrophysics, School of Physics and Astronomy, Rochester Institute of Technology, 84 Lomb Memorial Drive, Rochester, NY 14623, USA\label{aff17}
\and
Department of Physics and Astronomy, Texas A\&M University, College Station, TX, 77843-4242 USA\label{aff18}
\and
George P.\ and Cynthia Woods Mitchell Institute for Fundamental Physics and Astronomy, Texas A\&M University, College Station, TX, 77843-4242 USA\label{aff19}
\and
ESA/AURA Space Telescope Science Institute\label{aff20}
\and
Department of Physics and Astronomy, University of Kansas, Lawrence, KS 66045, USA\label{aff21}
\and 
Department of Astronomy, Indiana University, 727 East Third Street, Bloomington, IN 47405, USA\label{aff22}
}

   \date{Received ; accepted }

% \abstract{}{}{}{}{} 
% 5 {} token are mandatory
 
  \abstract
  {The dust content of star-forming galaxies is generally positively correlated with their stellar mass. However, some recent \textit{James Webb} Space Telescope (JWST) studies have shown the existence of a population of dwarf galaxies with an unexpectedly large dust attenuation. Using the Cosmic Evolution Early Release Science Survey (CEERS) data, we identified a sample of 1361 highly extincted low-mass (HELM) galaxies, defined as dwarf galaxies ($M_*\leq10^{8.5}$) with $A_V>1$~mag or more massive galaxies with an exceptionally high dust attenuation given their stellar mass (i.e., $A_V>1.6\,{\rm log_{10}}(M_*/{\rm M_{\odot}})-12.6$). The selection is performed using the multiparameter distribution obtained through a comprehensive spectral energy distribution (SED) fitting analysis, based on optical to near-infrared data. After excluding possible contaminants, like brown dwarfs, little red dots, high-$z$ ($z>8.5$) and ultra-high-$z$ ($z>15$) galaxies, the sample mainly includes sources at $z<1$, with a tail extending up to $z=7.2$. The sample has a median stellar mass of $10^7\,\rm M_{\odot}$ and a median dust attenuation of $A_V=2$~mag. We analysed the morphology, environment and star-formation-rate of these sources to investigate the reason behind their large dust attenuation. In particular, HELM sources have sizes (effective radii, $R_e$) similar to non-dusty dwarf galaxies and no correlation is visible between the axis ratios ($b/a$) and the dust attenuation. %For the two most numerous HELM subsamples we can exclude with a confidence of 89\% that HELM sources are more compact than non-dusty dwarfs. 
  This findings indicate that it is unlikely that the large dust attenuation is due to projection effects, but a prolate or a disk-on oblate geometry are still possible, at least for a subsample of the sources. We have found that the distribution of HELM sources is slightly skewed toward more clustered environments than non-dusty dwarfs and tend to be slightly less star forming. This finding, if confirmed by spectroscopic follow-up, indicates that HELM sources could be going through some environmental processes, such as galaxy interactions. }
  % % context heading (optional)
  % % {} leave it empty if necessary  
  %  {}
  % % aims heading (mandatory)
  %  {}
  % % methods heading (mandatory)
  %  {}
  % % results heading (mandatory)
  %  {}
  % % conclusions heading (optional), leave it empty if necessary 
  %  {}


   \maketitle
%
%-------------------------------------------------------------------

\section{Introduction}
Dust is a key and ubiquitous component of the interstellar medium (ISM), despite accounting for only $\sim1\%$ of its total mass \citep[see][]{Draine2003}. Indeed, it is central to the formation and cooling of the molecular gas and, therefore, to the formation of new stars. Moreover, dust profoundly impacts our understanding of galaxy evolution, given that it absorbs ultra-violet (UV) and optical radiation coming from stars re-emitting it at infrared (IR) wavelengths, impacting the observed spectra and the derived physical properties. Therefore, the dust content and its evolution with cosmic time remain a key topic in extragalactic astrophysics. \par

Dusty galaxies have been extensively studied from the local Universe to the early cosmic epochs, showing that these systems seem to dominate the cosmic star formation budget at least out to $z=4-5$ \citep{Zavala2017,Magnelli2024,Traina2025}. The existence and importance of deeply obscured objects, which are extremely faint or completely undetected in the rest-frame optical and UV, have been pointed out by several studies \citep[e.g.][]{Simpson2014,Franco2018,Wang2019,Talia2021,Gruppioni2020,Fudamoto2021,Enia2022,Gentile2024}. In recent years, the launch of the \textit{James Webb Space Telescope} (JWST) has allowed observing the stellar continuum of these objects in the near-IR to deeply investigate their physical properties. Before its advent, the majority of dusty galaxies were massive/luminous systems, but their stellar continuum was poorly known or impossible to study having only far-IR or radio observations. Now, with JWST, we are not only investigating the stellar continuum of the brightest dusty galaxies, but we are also pushing their redshift and stellar mass frontiers \citep[e.g][]{PerezGonzalez2023,Rodighiero2023,Williams2024,Gentile2025}. \par

It has been observed that the dust attenuation (\av) of star-forming galaxies has a positive correlation with stellar mass. On the one hand, more massive star-forming galaxies are generally more dusty and are dominated by sources detected in the mid-IR \citep[e.g., \textit{Spitzer};][]{PerezGonzales2005,PerezGonzales2008} and far-IR \citep[\textit{Herschel};][]{Rodighiero2010,Barro2019,Merida2023}. On the other hand, low-mass star-forming galaxies (with $M_{*}<10^{8.5}\,\rm M_{\odot}$) have a negligible dust content \citep{Pannella2015,McLure2018,Shapley2023,Liu2024}. This relation is shaped by the many complex mechanisms responsible for the creation and destruction of dust \citep[see][]{Draine2009,Salim2020,Schneider2024}. First, dust is produced by mass loss from asymptotic giant branch stars and supernova ejecta \citep[e.g.,][]{Gall2011,Sarangi2018}. Therefore, the correlation between dust attenuation and stellar mass can be seen as a byproduct of the correlation observed between stellar mass and star formation rate \citep[SFR;][]{Brinchmann2004,Noeske2007}. At the same time, dust in low-mass systems can be reduced due to stellar feedback-driven outflows, which can easily expel metal-enriched gas given their low gravitational potential \citep{Dayal2013}. Moreover, it has been argued that supernovae explosions could heat up the surrounding gas, reducing the amount of condensed dust \citep{Nozawa2011,Bocchio2016,Bianchi2007}. Finally, dust can also grow directly around ISM metals, with the growth that depend on local gas density, temperature and elemental abundance \citep{Draine1990,Mancini2015,Schneider2016,Popping2017,Graziani2020}.

All these phenomena that play a role in the formation of dust indicate a positive correlation between stellar mass and dust content and, as a consequence, dust attenuation. However, given the complexity and time scale of these mechanisms, we may also expect to observe galaxies deviating from such a relation. Indeed, \citet{Rodighiero2023}, \citet{Bisigello2023b} and \citet{Gandolfi2025} have revealed an unexpected population of low-mass galaxies with $M_{*}<10^{8.5}\,\rm M_{\odot}$ and a non-negligible dust attenuation (\av$>1$) mainly, but not only, located at $z<2$, by looking for galaxies detected at 4\mum but undetected at $\lambda\leq2$\mum. This population of highly extincted low-mass galaxies (HELM galaxies hereafter) is too faint in the optical and in the far-IR to be observed with past and current facilities, like the \textit{Hubble} (HST) or \textit{Herschel} space telescopes, but can now be explored in the near-IR with JWST. Moreover, some HELM galaxies, being faint and red, mimic the colour expected for ultra-high redshift galaxies $z>15$ \citep{Gandolfi2025,PerezGonzalez2025,Castellano2025}, making them a poorly characterised class of interlopers when studying the extremely high-$z$ Universe.

In this work, we take a step forward with respect to the pioneering studies by \citet{Bisigello2023b} and \citet{Rodighiero2023} to explore the nature of HELM galaxies using a larger and purer sample based on physical properties instead of colour selections. The paper is organised as follows. In Section \ref{sec:data} we describe the data used and sample selection. Given the faintness of HELM galaxies, we perform some preliminary analysis on the stacked photometry in Section \ref{sec:stack}, while we discuss different scenarios for the large dust attenuation of HELM galaxies in Section \ref{sec:nature}. Throughout the paper, we consider a $\Lambda$CDM cosmology with $H_0=70\,{\rm km}\,{\rm s}^{-1}\,{\rm Mpc}^{-1} $, $\Omega_{\rm m}=0.30$, $\Omega_\Lambda=0.70$, a Chabrier initial mass function \citep[IMF,][]{Chabrier2003}, and all magnitudes are in the AB system \citep{Oke1983}.

\section{Data} \label{sec:data}

In this work, we used data from the Cosmic Evolution Early Release Science Survey \citep[CEERS][]{Finkelstein2025}, which covered $\sim90\,\rm arcmin^2$ of the Extended Groth Strip (EGS) with both spectroscopy and photometry. In particular, we considered the public release of the imaging data (v.0.6) presented in detail in \citet{Bagley2023}, which include both observing epochs and six broad bands (F115W, F150W, F200W, F277W, F356W, F444W) and one medium-band Near Infrared Camera (NIRCam) observation (F410M). For $\sim70\%$ of the EGS field, there are also Mid-Infrared Instrument (MIRI) data with various filter combinations. On the one hand, the CEERS programme includes sparse MIRI coverage \citep{Yang2023}, with a filter combination that may include only two deep short-wavelength filters (F560W and F770W) or six broad-band filters covering 5.6\mum to 21\mum with a shallower depth, depending on the pointing. On the other hand, observations from 7.7\mum to 21\mum are also available as part of the MIRI EGS Galaxy and AGN \citep[MEGA][]{Backhaus2025}. However, the depths of MIRI observations are not enough to match the faintest galaxies detected with NIRCam. Deep HST observations are also available in the entire field of CANDELS \citep{Koekemoer2011,Grogin2011} in six filters (ACS/WFC F606W, F814W and WFC3/IR F105W, F125W, F140W, F160W) from a variety of different programmes and were all combined into mosaics at a scale of 30 mas/pixel.

Finally, we verified whether any of the sources of interest have available spectroscopic data, coming from the CEERS NIRCam grism observations and derived using a custom extraction and contamination modelling similar to \citet{Pirzkal2018}, or Near Infrared Spectrograph (NIRSpec) Multi-Object Spectroscopy observations with medium grating or the PRISM (Arrabal Haro et al., in prep), or from the DAWN JWST Archive (DJA\footnote{\href{https://dawn-cph.github.io/dja}{https://dawn-cph.github.io/dja}}). More details are given later for specific objects.

\subsection{Photometric catalogue} \label{sec:cat}
This work is based on the source catalogue described by \citet{Finkelstein2024} and based on the NIRCam mosaic. Briefly, short-wavelength HST and NIRCam filter images (F606W, F814W, F115W, F150W and F200W) were all matched to the point-spread function (PSF) of the F277W filter. At the same time, for bands with larger PSFs, like the redder NIRCam bands and all HST/WFC3 bands, they were corrected for missing flux using correction factors derived by convolving the F277W image to the larger PSF and measuring the flux ratio in the native image to that in the convolved image. Photometry was then performed with \texttt{Source Extractor} \citep[v2.25.0][]{Bertin1996} in two-image mode, using as the detection image the inverse-variance weighted combination of the native-resolution F277W and F356W band images. Fiducial fluxes were measured in small elliptical Kron apertures, but later corrected for missing flux \citep[see][for more details]{Finkelstein2023a}.

Given the sparsity and depth of MIRI data, we decided not to include the fiducial fluxes in the catalogue and the SED fitting, but to check a posteriori for the detection of single objects or in the stacked images. The final catalogue includes 101\,808 sources.

\subsection{SED fitting}\label{sec:SEDfitting}
We fit the 13 HST and NIRCam filter observations with the Bayesian Analysis of Galaxies for Physical Inference and Parameter EStimation \citep[\textsc{Bagpipes}][]{Carnall2018} code to the entire sample of 101\,808 galaxies extracted in Sec. \ref{sec:cat}. We include in the fit also filters with non detections, that is $S/N<3$, including directly the measured fluxes and uncertainties in order to include all available photometric information without imposing strong upper limits. We added in quadrature 5\% of the flux to the flux uncertainties to avoid small errors to dominate the fit and to include unknown uncertainties, like uncertainties in SED modelling. 

We considered the 2016 version of the \citet{Bruzual2003} single stellar population models with metallicity ranging from $5\%$ solar to 2.5 times solar. As a first run, we considered a delayed star-formation-history (SFH) model $\propto t\,e^{-(t/\tau)}$, with $\tau$ varying between 10\,Myr and 10\,Gyr and the age $t$ ranging from 1\,Myr to 15\,Gyr (or the maximum age of the Universe at a given redshift, automatically included in the code). The redshift ranges from $z=0$ to $z=15$, with a flat uninformative prior, while stellar masses are allowed to vary between $10^{6}$ and $10^{12.5}\,M_{\odot}$ considering a flat and uninformative prior. We included nebular continuum and emission lines from a precomputed model grid considering an ionisation parameter (i.e., dimensionless ratio of densities of ionising photons to hydrogen) between $\rm log_{10}(U)=-4$ and $-2$. Finally, for dust attenuation, we considered the reddening law by \citet{Calzetti2000}, leaving the optical dust attenuation free to vary between 0 and 6 magnitudes. In addition, we performed an additional run that forced the dust attenuation to be below $A_V=1\,$mag.\\
We also performed three other runs, varying the SFH or the dust attenuation law and leaving all the other parameters unchanged. In particular, first we considered the Small Magellanic Cloud (SMC) attenuation curve \citep{Gordon2003}, varying the $A_V$ between 0 and 6 magnitudes and keeping a delayed SFH. Second, we kept the reddening law by \citet{Calzetti2000} and changed the SFH to an exponentially declining one (that is, $\propto \,e^{-(t/\tau)}$), with $\tau$ varying between 10\,Myr and 10\,Gyr. Finally, we changed the SFH to a double power law (that is, $\propto \,[(t/\tau)^{\alpha}+(t/\tau)^{-\beta}]^{-1}$), with $\tau$ varying between 10\,Myr and 10\,Gyr and the two power law exponents $\alpha$ and $\beta$ ranging from 0.01 to 1000, with logarithmic priors.

\subsection{Sample selection}\label{sec:selection}

\begin{table}[]
    \caption{Summary of the steps considered for the sample selection.}
    \resizebox{\columnwidth}{!}{
    \centering
    \begin{tabular}{c|ccc}
    Selection  & \multicolumn{3}{c}{Number of objects} \\
         & HELM-$1\sigma$ &  HELM-$2\sigma$ &  HELM-$3\sigma$\\
         \hline
        Initial HELM selection$^{a}$ & 3955 & 203 & 19 \\
        \hline
        Brown-dwarf$^{b}$ & 3 & 0 & 0 \\
        $\rm \chi^{2}(Av<1)<\chi^{2}(Av\,free)$ & 1326 & 28 & 1 \\
        $z>8.5^{c}$ & 3 & 1 & 0 \\
        Spectroscopic data$^{d}$ & 9 & 1 & 1 \\
        Visual inspection & 1425 & 15 & 3\\
        \hline
        Final HELM sample & 1189 & 158 & 14\\
    \end{tabular}
    \label{tab:selections}}
    \tablefoot{$^{a}$ See eq. \ref{eq:selection}; $^{b}$ matching with the sample by \citet{Holwerda2024}; $^{c}$ matching with the sample by \citet{Finkelstein2024}; $^{d}$ sources that are removed from the HELM sample thanks to spectroscopic data (Sect. \ref{sec:spec}). }
\end{table}

\begin{figure}
    \centering
    \includegraphics[width=\linewidth,keepaspectratio]{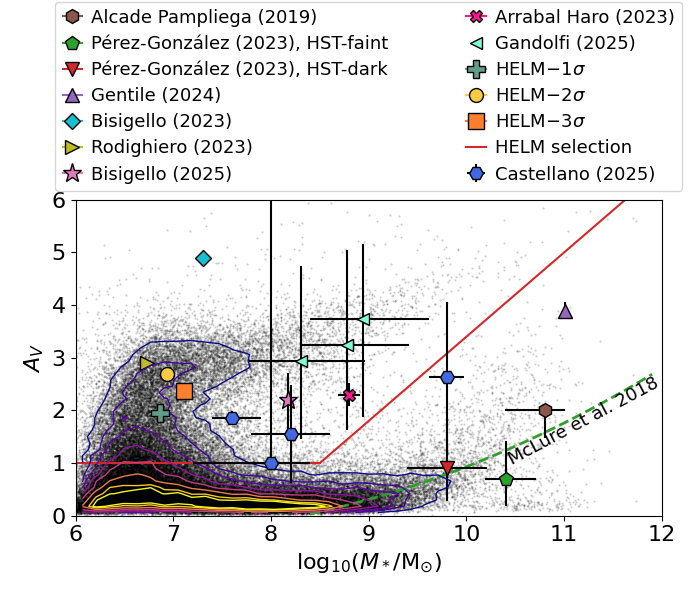}
    \caption{The unexpected population of HELM galaxies, as visible in the stellar mass and dust attenuation plane. Black dots are the entire CEERS sample (before any selection), as derived from the first SED fitting run described in Sec. \ref{sec:SEDfitting} and considering the median value of both Av and stellar mass. Contour lines show the 20 (yellow) to 90\% (dark blue) of the distribution, in steps of 10\%. Coloured symbols show other samples of dusty objects from the literature \citep{AlcaldePampliega2019,PerezGonzalez2023,Gentile2024,Rodighiero2023,Bisigello2023b,Gandolfi2025,Castellano2025,ArrabalHaro2023b} and the one analysed in this paper (see Sec. \ref{sec:selection}). \citet{Bisigello2025a} is an HELM galaxy confirmed spectroscopically and is also part of the HELM-$2\sigma$ sample, while the source by \citet{Castellano2025} has been confirmed spectroscopically, but it is on another field. The green dashed line shows the known relation between dust attenuation and stellar mass \citep{McLure2018} derived at $2<z<3$, while the red solid line shows the selection criterion of HELM sources.}\label{fig:MAv}
\end{figure}

From the results of the previously mentioned \textsc{Bagpipes} runs, we removed objects with a bad fit (i.e., $\chi^{2}>20$, 13398 objects) and objects below the $5\sigma$ detection limit in the F444W filter \citep[i.e., 28.7~mag,][34795 sources]{Finkelstein2025}. We consider the F444W for a detection given the red nature of the HELM sources. Looking at the stellar mass vs. dust absorption plane for the all CEERS sample (Fig. \ref{fig:MAv}), we selected HELM galaxies as objects with:
\begin{align}\label{eq:selection}
    &{\rm log_{10}}(M_*/{\rm M_{\odot}})\leq8.5\land A_V>1 \;\lor\\
    &{\rm log_{10}}(M_*/{\rm M_{\odot}})>8.5 \land A_V>1.6\,{\rm log_{10}}(M_*/{\rm M_{\odot}})-12.6.\nonumber
\end{align}
using the complete multiparameter distributions of all SED fitting runs. This selection was chosen to separate all objects with a dust attenuation above the expectation given their stellar mass. Therefore, we selected dwarf galaxies with some dust attenuation (first criterion), but also more massive galaxies with an extreme dust attenuation (second criterion).

With the above selection, we identified 3955 plausible, 203 robust, and 19 very robust HELM candidates, with 68\% to 95\% (1$\sigma$), 95\% to 99.7\% (2$\sigma$) and more than 99.7\% (3$\sigma$) of the output multiparameter distributions within the HELM selection, respectively. In the following, we call these sub-samples HELM-1$\sigma$, HELM-2$\sigma$, and HELM-3$\sigma$. These percentages are derived averaging the percentages of the four single SED fitting runs and therefore indicate how frequent is the HELM solution among all other possibilities. For example, the minimum percentage of the multiparameter distribution that is within the HELM selection for a single SED fitting run is 98.8\% for the HELM-3$\sigma$ sample and 80\% for the HELM-2$\sigma$ sample. On the other hand, one of the four SED runs could be completely outside the HELM selection for the HELM-1$\sigma$ sources, as long as the others are inside for $>90\%$.

We cross-matched these catalogues with the sample of brown dwarfs by \citet{Holwerda2024}, which is derived using the same dataset. We found three objects in the HELM-1$\sigma$ with a probability higher than 0.7 to be a brown dwarf \citep[see][for the details on the classification]{Holwerda2024}. We removed these brown dwarf candidates from the catalogue, reducing the sample size to 3952 objects. 

We then removed objects for which the fit forcing $A_V<1\,$mag has a lower $\chi^{2}$ than the run with a free-to-vary dust attenuation. For this comparison we considered the results obtained with the delayed SFH. We remind the reader that the HELM selection in Eq. \ref{eq:selection} is based on the multidimensional probability distribution and not on the best point estimates. Therefore, this additional criterion verifies that the HELM solution is not only the most represented in the multidimensional probability distribution, but also the one with the lowest $\chi^{2}$. With this additional criterion we removed one object from the HELM-3$\sigma$ sample, while for all the other objects in the same sample the difference in $\chi^{2}$ is $\Delta\chi^{2}>1.6$, with a median value of $\Delta\chi^{2}=6.6$. In the HELM-$2\sigma$ sample, we removed 28 objects for which the $A_V<1\,$mag solution is preferred, suggesting the possibility that their red nature is not due to reddening. Around one third of the HELM-$1\sigma$ sample has a non-dusty solution with a lower $\chi^{2}$. This is in line with the fact that these sources are the less robust HELM candidates. The HELM-$1\sigma$ sample is therefore reduced to 2626 objects.

We also cross-matched all HELM samples with the sample of $z>8.5$ galaxy candidates by \citet{Finkelstein2024}. There are no matches with the HELM-$3\sigma$, while there is one with the HELM-$2\sigma$, and three with the HELM-$1\sigma$. On the one hand, the source in the HELM-$2\sigma$ (CEERS\_90671) has a spectroscopic redshift of $z=8.638$ \citet{ArrabalHaro2023a}, confirming its high-$z$ nature. The wrong HELM identification is possibly link to the degeneracy between redshift and dust attenuation, which is strengthened by the relatively faintness of this source ($F277W=$28.1 mag) and the presence of some dust attenuation $A_V=$0.2--0.9 \citep{ArrabalHaro2023a}.
On the other hand, there is no spectroscopic confirmation for the 10 candidate high-$z$ sources in the HELM-$1\sigma$ sample. Given that our purpose is to obtain a sample as pure as possible, we decided to remove these high-$z$ candidates from our samples anyway, but we performed further comparisons to investigate if there are any other possible high-$z$ candidates in Sect. \ref{sec:highz}. In addition, there was no overlap, by construction, between the HELM samples presented here and the $z>15$ catalogue by \citet{Gandolfi2025}. We also verified that there is no overlap with the $z>15$ catalogue by \citet{Castellano2025}, as these sources had less than 20\% of their multiparameter distribution inside the HELM selection (see Section \ref{sec:selection}).

We also verified the availability of spectroscopic data to confirm or disclaim the HELM nature of our sources. The objects with spectroscopic data are discussed in detail in the Appendix \ref{sec:spec}, but the analysis allowed us to remove one object from the HELM-$3\sigma$ sample, one from the HELM-$2\sigma$ sample and nine from the HELM-$1\sigma$, while one HELM source from the HELM-$2\sigma$ was confirmed \citep{Bisigello2025a}. The spectroscopic data were not sufficient to confirm or reject the HELM solution for other two HELM-$1\sigma$ sources, as no evident features were observed in the spectra. We highlight that these spectroscopic follow-ups were fine tuned to high-$z$ or massive galaxies, so an unbiased spectroscopic follow-ups of the HELM galaxies is necessary to verify the level of contamination. The spectroscopically confirmed HELM source CEERS-93316 \citet{ArrabalHaro2023a}, which is at $z=4.912\pm0.001$, has been removed in our selection procedure having $\chi^2>20$. In this paper we aim to a selection as pure as possible, at the expense of completeness.

\begin{figure}
    \centering
    \includegraphics[width=\linewidth,keepaspectratio]{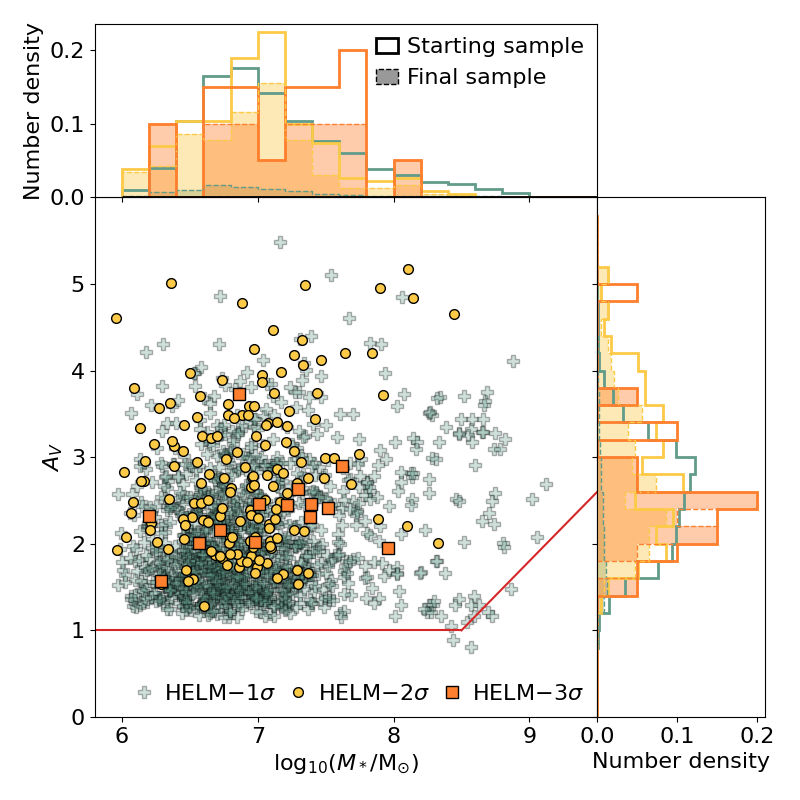}
    \caption{Stellar mass and dust attenuation plane showing the HELM sources and the HELM selection criterion (red solid line). For the stellar mass and dust attenuation we show the median value of the median outputs of each SED setup. We remind that the HELM selection is performed on the multiparameter probability distribution and not on the point estimates of the physical properties, so the median value shown here can be slightly out of the selection if one of the SED runs has a median output distant from the HELM selection. The two side panels show the number density of the HELM samples before and after the selections of Table \ref{tab:selections}.}\label{fig:MAv_HELM}
\end{figure}

As a final step, we visually inspected all remaining sources to remove artifacts, which are more frequent in the HELM-$1\sigma$ sample. Given that many sources are detected only in a few bands, imposing a low $\chi^2$ does not guarantee that the source is real and without photometric issues. In Appendix \ref{sec:vis} we show some example of the objects we removed, which include sources with strange shapes, sources at the edge of images or outside the field-of-view in some filters, but also cases in which no source is visible at the expected position, due to issues in the extraction procedure.
The final HELM samples consist of 1189 plausible sources in the HELM-$1\sigma$, 158 robust objects in the HELM-$2\sigma$, and 14 very robust candidates in the HELM-$3\sigma$ sample. A summary of the selection steps is reported in Table \ref{tab:selections}. The stellar mass and dust attenuation (median values of the median outputs of all SED runs) for all the HELM sources is shown in Fig. \ref{fig:MAv_HELM}. Of all these sources, only two HELM-$1\sigma$ galaxies are detected in MIRI. One source, which is part of the CEERS survey \citep{Yang2023}, is detected at 12.8\mum with a $S/N=7.7$ and
undetected at 18\mum. The other source has been observed in the MEGA survey \citep{Backhaus2025}. It has a $S/N>40$ in the 7.7\mum filter, but is undetected ($S/N<1$) in the other bands at $\lambda=10$, 15 and 21\mum. However, the first object has a suspicious annulus shape, indicating that there is probably an artifact in the F1280W image, while the second is close to the edge of the MIRI images. Therefore, neither object can be identified as a secure MIRI detection.

We also identified a control sample of low-$z$ non-dusty dwarf galaxies to compare their physical properties with the ones of HELM galaxies to investigate the reasons of their large dust attenuation. This control sample was selected as galaxies having a good fit (i.e., $\chi^{2}<20$), a median redshift $z<1$, negligible dust attenuation $A_V<1$ and low stellar mass ${\rm log}_{10}(M_{*}/\rm M_{\odot})\leq8.5$, considering a confidence of $84\%$ on the first SED fitting run (i.e., delayed SFH and \citet{Calzetti2000} reddening law). The control sample consists of 15\,045 galaxies, with $A_{V}=0.29\pm0.14$ and ${\rm log}_{10}(M_{*}/\rm M_{\odot})=6.8\pm0.6$ on average.

\subsection{Comparison between HELM samples}
In Fig. \ref{fig:MAv} we report the dust attenuation and stellar mass derived for the entire CEERS sample considering the first SED fitting run (i.e., delayed SFH, Calzetti's reddening law) and the HELM selection criterion. From this plot, it is evident how the two-dimensional distribution show a bimodality, with the majority of galaxies following the expected relation between stellar mass and dust attenuation, while other (forming our HELM samples) are well above the expectation.
In the same Figure we also report, for comparison, the median stellar mass and dust attenuation of the three HELM samples analysed in this work and other samples of dusty sources from the literature \citep{AlcaldePampliega2019,PerezGonzalez2023,Gentile2024,Rodighiero2023,Bisigello2023b,Castellano2025,ArrabalHaro2023b}. The first four samples were selected for being faint or dark in the optical, but bright in the near-IR or at radio frequencies and are more massive than the bulk of the HELM population. The sample presented in \citet{Bisigello2023b} is generally more dust-obscured than our HELM sample. It is based on the first epoch of the CEERS survey, but it was selected for having $\rm S/N>3$ at 4.4\mum and  $\rm S/N<2$ at $\lambda<2$\mum. \citet{Rodighiero2023} considered a similar sample selection, but it is based on deeper data, which results in a sample with dust attenuation similar to the bulk of our HELM galaxies, but slightly less massive. The sample by \citet{Gandolfi2025} consists of three HELM candidates which however have some probability of being at $z>15$. The five galaxies presented in \citet{Bisigello2025a}, \citet{ArrabalHaro2023b} and \citet{Castellano2025} are the only spectroscopically confirmed HELM galaxies.

\begin{figure}
    \centering
    \includegraphics[width=\linewidth]{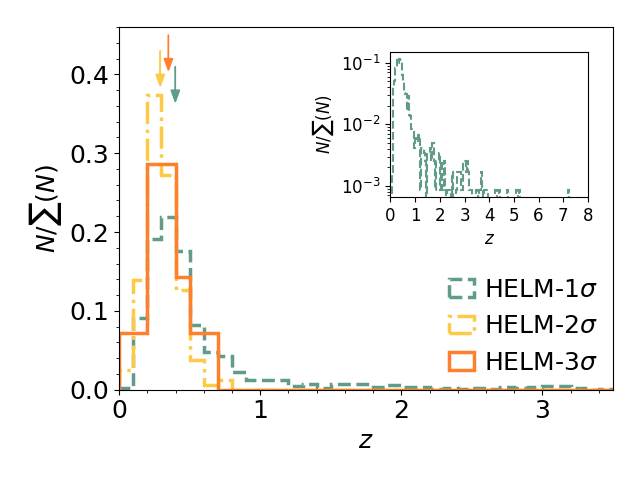}
    \includegraphics[width=\linewidth]{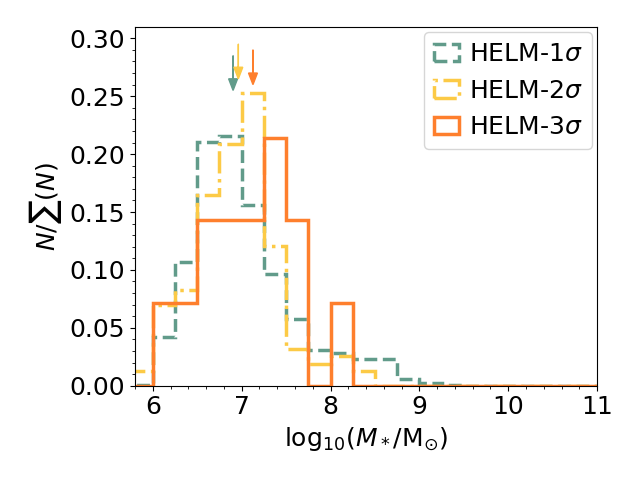}
    \caption{Normalized redshift (top) and stellar mass (bottom) distributions of the HELM samples. Coloured arrows indicate the median of the three samples and are slightly shifted vertically for clarity. In the top panel, we limit the plot to $z<3.5$ for visual clarity, but we show the entire range in the inset for the HELM-$1\sigma$ sample.}
    \label{fig:zM_dist}
\end{figure}

We verified that there are no strong biases between the different SED fitting runs for the HELM samples (see Appendix \ref{sec:sedcomparison}). The redshift and stellar mass distributions of the HELM samples are shown in Figure \ref{fig:zM_dist}. The HELM samples present similar redshift distributions, with a median value ranging from $z=0.29$ to $z=0.40$. However, all galaxies in the HELM-$3\sigma$ sample are at $z\leq0.6$, while the HELM-$2\sigma$ sample presents one source at $z>1$, with a spectroscopic redshift of $z=4.883$ \citep{Bisigello2025a}. The HELM-$1\sigma$ sample instead extends up to $z=7.2$. Therefore, the large majority (90\%) of HELM sources are at $z<1$. As for the stellar mass, all samples have median values between ${\rm log}_{10}(M_{*}/\rm M_{\odot})=6.9$ and ${\rm log}_{10}(M_{*}/\rm M_{\odot})=7.2$, but the HELM-$1\sigma$ presents a tail at ${\rm log}_{10}(M_{*}/\rm M_{\odot})>8.5$ that corresponds to the galaxies at the highest redshift. We performed the two-sided Kolmogorov-Smirnov (KS) test to exclude the possibility that the HELM samples are extracted from the same distributions. We can exclude with $90\%$ (80\%) confidence that the HELM-$1\sigma$ sample is extracted from the same redshift distribution as the HELM-$2\sigma$ (HELM-$3\sigma$) sample. We cannot exclude the possibility that the HELM-$2\sigma$ and HELM-$3\sigma$ samples are drawn from the same redshift distribution, or that any of the three HELM samples originate from the same stellar mass distribution.

Given the redshift distributions of the samples, we will concentrate our analysis at $z<1$, corresponding to 1053/1189 of the HELM-$1\sigma$ sample, 157/158 of the HELM-$2\sigma$ sample, and all 14 objects of the HELM-$3\sigma$ sample. 

\begin{figure}
    \centering
    \includegraphics[width=\linewidth,trim={20 20 20 10},clip]{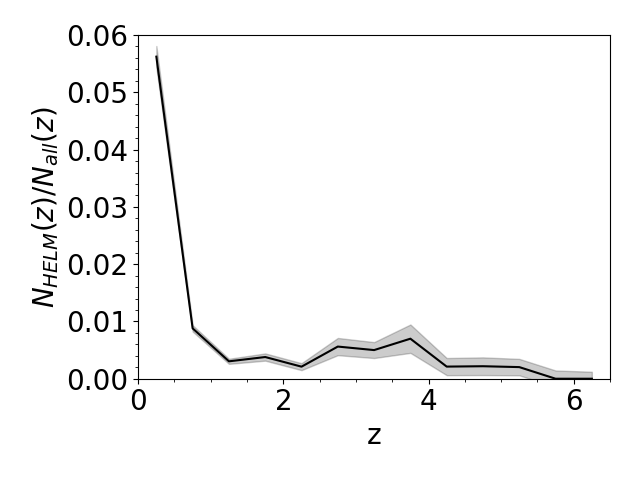}
    \includegraphics[width=\linewidth,trim={25 20 20 10},clip]{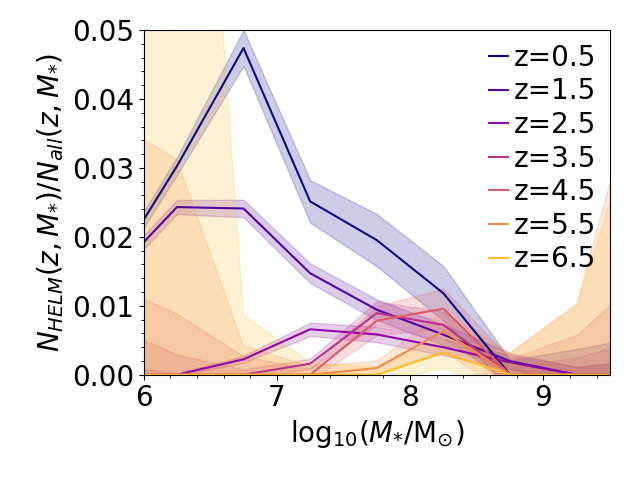}
    \caption{Fraction of HELM sources, considering all samples, with respect to all galaxies with similar redshift (top) and similar redshift and stellar mass (bottom). We consider only galaxies in the entire CEERS sample with good SED fit (i.e., $\chi^2<20$).}
    \label{fig:fhelm}
\end{figure}

It is worth noticing that HELM galaxies overall represent the minority with respect to the whole sample of observed galaxies, as they are 1.5$\%$ of the entire CEERS sample with a good SED fit. Their faint nature and their rarity, maybe linked to a short-lived evolutionary phase, are probably the reason behind their absence in previous studies of dusty objects. They are also a minority among galaxies with similar stellar mass and redshift (Fig. \ref{fig:fhelm}), therefore their IR flux would be missed even in stacking analysis based on stellar mass selections. Indeed, HELM sources account at maximum for less than 6\% of all sources at $z<1$ and with stellar mass around $\rm10^7\,M_{\odot}$, while their fraction decreases at higher stellar masses and redshifts. For example, HELM sources are $<1\%$ of all the galaxies at $z>1$. It is however necessary to consider that we did not correct for completeness and we expect the sample of HELM sources to be more incomplete than non-dusty dwarfs, considering their faintness at short wavelengths.

\section{Stacked photometry} \label{sec:stack}
As a first step, we perform a stacking analysis on the photometry of the HELM samples to verify their red nature. Indeed, the faint nature of these galaxies makes them undetected in many bands, as, on average, sources in the HELM-$1\sigma$, HELM-$2\sigma$, and HELM-$3\sigma$ have $S/N>3$ in 4.5, 6.8, and 8.8 filters, respectively, out of the 13 HST and NIRCam bands available, but also no detections in the MIRI bands. The stack is performed in the rest-frame, combining all three samples and splitting them in redshift bins up to $z=1.1$. \par

We started by verifying that the source is inside the observation footprint of the considered band, as it varies from filter to filter. Then, we extracted an area of $3\arcsec \times 3\arcsec$ centred on each source. We stacked 11 objects at $z<0.11$, 206 at $0.11\leq z<0.24$, 414 at $0.24\leq z<0.38$, 341 at $0.38\leq z<0.53$, 140 at $0.53\leq z<0.71$, 66 at $0.71\leq z<0.90$, and 28 at $0.90\leq z<1.11$. Redshift bins are chosen considering the rest-frame wavelength traced by each filter at different redshifts. The same stacking procedure is also applied to the control sample of not-dusty dwarfs. 

\subsection{Background removal}
Before performing the stack of each object, we removed any residual local background, following the procedure presented in \citet{Rodighiero2023}.
In particular, we modelled the 1D pixel intensity distribution with the sum of a Gaussian function, whose mean corresponds to the background, and a Schechter function, that is a minor component accounting for sources within this region. The fit is performed using as initial values the maximum, the median and 70\% of standard deviation of the 1D pixel intensity distribution for the amplitude, mean and standard deviation of the Gaussian function. For the Schechter function we considered as initial values an amplitude of 10\% the maximum of the 1D pixel intensity distribution for the amplitude, a knee equal to the standard deviation of the 1D pixel intensity distribution and a slope of 2. The fit is stable against reasonable changes in these initial values.

\subsection{Photometric extraction and comparison}
We normalize each object by its flux in the F356W filter and then we combined all images and we derived the median pixel-by-pixel, after performing a $3\sigma$ clipping. We extracted the photometry associated with each cutout using a $0\arcsecf3$ aperture radius, which is larger than the stacked sources. Uncertainties on the photometry were derived from the rms of the flux in four empty regions with a $0\arcsecf3$ aperture radius located at the corners of each stacked image. \par 

\begin{figure}
    \centering
    \includegraphics[width=\linewidth,trim={20 50 20 10},clip]{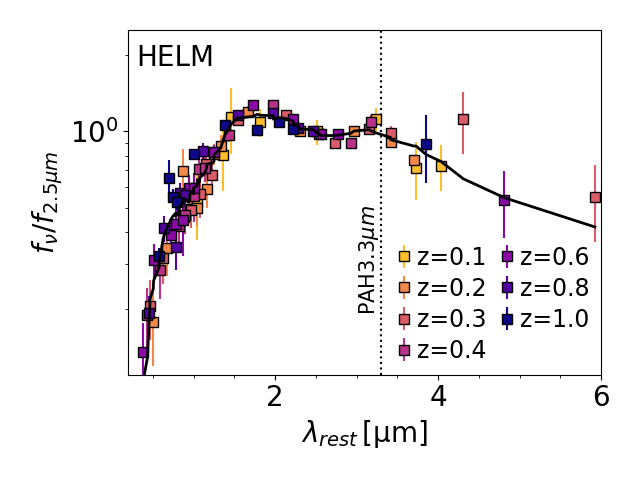}
    \includegraphics[width=\linewidth,trim={20 20 20 10},clip]{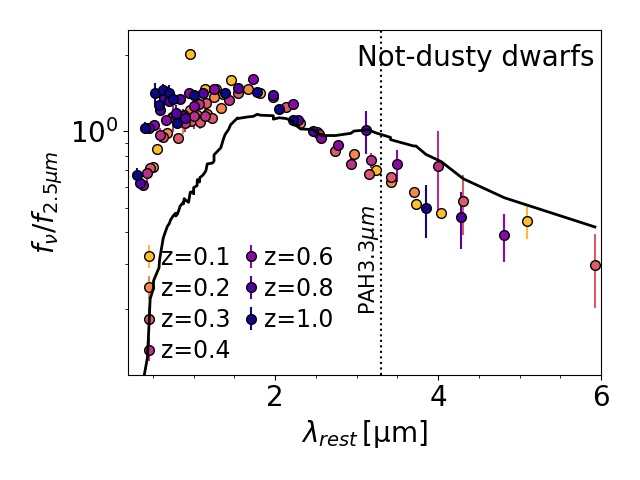}
    \caption{Comparison between the stacked rest-frame photometry of the combined HELM samples (top) and the sample of non-dusty dwarfs (bottom) at different redshifts. Stacked photometry is normalised at $2.5\,\mu m$. We show the HELM SED (solid black lines), convolved using a top-hat function over seven data points, in both panels for comparison. }
    \label{fig:stack}
\end{figure}

In Figure \ref{fig:stack} we report a comparison between the photometry of the HELM sample and the control sample, all normalised to the flux observed at rest-frame $2.5\,\rm\mu m$. The coherent behaviour of both stacks support the validity of the majority of the photometric redshifts, despite the faint nature of these objects. Some discrepancies are visible in the short-wavelengths slopes of both stacks. However it is evident that the HELM galaxies have not only a redder continuum at $\lambda<2\,\mu m$, but they show also a flatter continuum at longer wavelengths. Moreover, there is a tentative emission feature right at the position of the $3.3\mu m$ polycyclic aromatic hydrocarbon (PAH) feature. We performed a linear fit of the continuum blueward and redward the tentative PAH feature to derive an estimate of its equivalent width (EW). In particular, we considered the excess with respect of the continuum from $z=0.1$ to $z=0.3$, when the tentative feature is traced by the F356W, F410M, and F444W filters. The EW varies from $298\pm5\,$nm ($z=0.2$) to $659\pm6\,$nm ($z=0.3$). Spectroscopic follow-ups are fundamental to verify if indeed the $3.3\mu m$ is present and so prominent.

\section{On the nature of HELM galaxies} \label{sec:nature}
In this section we investigate some possible scenarios to explain the high dust attenuation observed in HELM galaxies. In these comparisons, except when analysing colours, we limit the analysis to objects at $z<1$, as the limited number of sources at higher redshifts precludes the study of any redshift evolution. 

%%%%%%%%%%%%%%%%
\subsection{Are we misinterpreting ultra-high-$z$ galaxies?}\label{sec:highz}

\begin{figure}
    \centering
    \includegraphics[width=\linewidth,trim={23 25 22 20},clip]{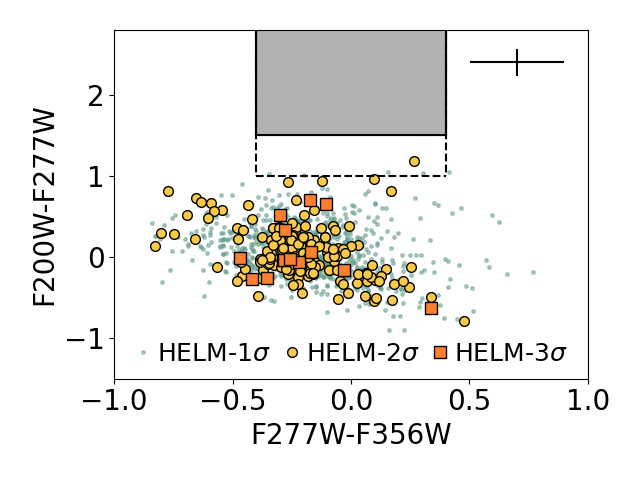}
    \includegraphics[width=\linewidth,trim={23 25 7 20},clip]{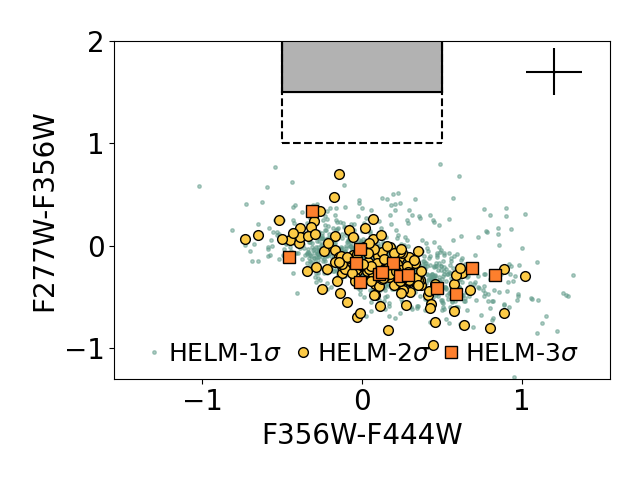}
    \caption{Colour-colour F277W-F356W vs. F200W-F277W (top) and
    F277W-F356W vs. F356W-F444W (bottom) diagram of
the HELM samples. The grey shaded area and the black dashed lines are the golden and silver selection criteria for $z=15-20$ (top) and $z=20-30$ (bottom) by \citet{Castellano2025}. We show only HELM galaxies with $\rm S/N > 3$ in the considered filters. Median errors are reported in the top right of the plot. }
    \label{fig:z20}
\end{figure}

As highlighted by several works \citep[e.g.,][]{Bisigello2023,Zavala2023,Gandolfi2025,Castellano2025}, dusty galaxies at $z<6$ can be misinterpreted as ultra-high-redshift sources beyond $z>15$, particularly when they are highly star-forming. To test whether this is a possibility for our HELM sources, in Fig. \ref{fig:z20} we report the selection criteria for galaxies at $z=15-20$ and $z=20-30$ by \citet{Castellano2025}. None of our sources are inside the more strict selection criteria for $z=15-20$ and $z=20-30$ galaxies (shaded region), while one HELM-2$\sigma$ source and four HELM-1$\sigma$ galaxies are inside the less conservative selection (dashed black line). However, it is necessary to consider that our sources have $S/N>2$ in bands that should be blueward of the Lyman-break, showing that they are likely not ultra-high-z sources. One HELM-1$\sigma$ galaxy is instead at the edge of the less conservative selection criteria for $z=20-30$ selection. All our HELM-$3\sigma$ sources are outside the selection criteria, but the large photometric errors could make some of our sources scatter inside the colour selection.
In this regard, future spectroscopic follow-ups are fundamental to distinguish between HELM and $z>15$ sources and further improve their selection.

From the same figure it is evident that sources that are red in the F200W-F277W colour are blue in F277W-F356W and red in F356W-F444W (and viceversa). This behaviour is caused by the tentative 3.3$\,\mu m$ PAH feature, as visible from the stacked photometry in Fig. \ref{fig:stack}.
%%%%%%%%%%%%%%%%

\subsection{Are we misinterpreting passive galaxies?}
There is a well known degeracy between dust attenuation and stellar age, therefore we investigate the possibility that we are misinterpreting passive galaxies as sources with large dust attenuation. To do so, we performed an additional SED fitting run where we forced a passive soultion using \textsc{Bagpipes}. In particular, we have considered a burst of star formation with an age range between 1 Gyr and the age of the Universe at the fitted redshift. The remaining parameters are the same as the main run described in Section \ref{sec:SEDfitting}. After this run, we have that all HELM-$3\sigma$ sources, 156 out 158 (99\%) HELM-$2\sigma$ galaxies, and 1105 out of 1189 (93\%) HELM-$1\sigma$ objects are still selected as HELM sources, even if they are completely passive. This suggest that these sources are too red to be described simply as passive galaxies and they require anyway a large amount of dust attenuation.

%%%%%%%%%%%
\subsection{Are HELM galaxies faint little red dots?}
In the past couple of years JWST has revealed a new population of red, compact sources possibly hosting an active galactic nuclei and mainly present at $z>4$ \citep[e.g.][]{Kocevski2023,Kocevski2024,Harikane2023,Matthee2024,Greene2024,Labbe2023,Labbe2024,Killi2024,Furtak2023,Taylor2024,Durodola2024,PerezGonzalez2024}, but with a possible sub-population extending at lower redshifts \citep{Q1-SP011,Ma2025}. The nature of these sources, called "little red dots" (LRDs), is still highly uncertain, but due to their red colour, faintness, and compactness, they may partially overlap with HELM galaxies. In particular, LRDs constitute around $1\%$ of the population of galaxies with $M_{UV}\sim-19$ at $z\sim5$ \citep[see, e.g.,][]{Kocevski2024}, which is similar to the percentage of HELM galaxies at lower redshift. Here we want to verify whether we are misinterpreting our sources, verifying if they could instead be the faint tail of $z>4$ LRDs.

\begin{figure}
    \centering
    \includegraphics[width=\linewidth,trim={23 25 7 20},clip]{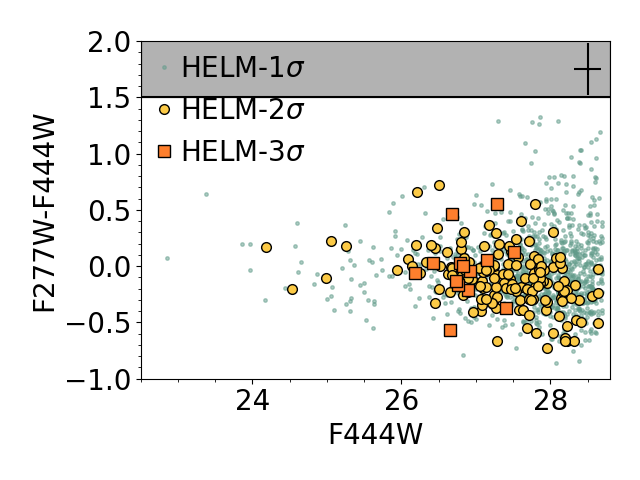}
    \includegraphics[width=\linewidth,trim={23 25 7 20},clip]{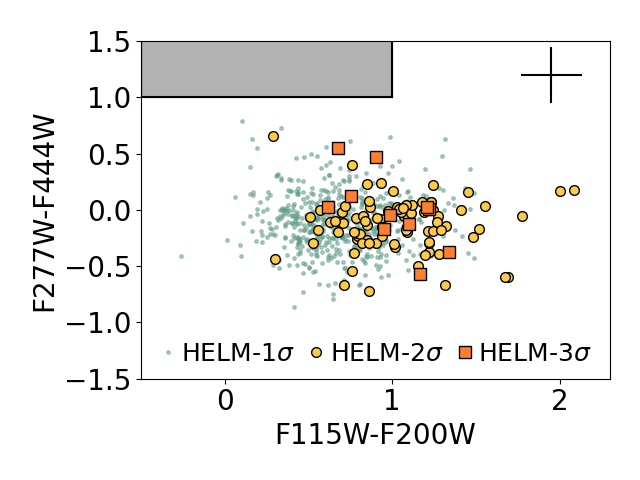}
    \caption{Colour-magnitude F444W-F277W vs. F444W magnitude (top) and colour-colour F444W-F277W vs. F200W-F115W diagram (bottom) of the HELM samples. The grey shaded area is the LRD selection criteria by \citet[][top]{Barro2024} and by \citet[][bottom]{Greene2024}. We show only HELM galaxies with $S/N>3$ in the considered filters. The median errors are shown in the top right.}
    \label{fig:LRD}
\end{figure}

First, we compared our samples with the sample of LRDs identified by \citet{Kocevski2023} based on the rest-frame UV and optical slopes to identify objects with the "v-shape" SED characteristic of LRDs. This sample is derived from the same CEERS dataset used in this work and we found no source in common. Second, we considered the possibility that our sources are so faint that we do not observe the blue UV rest-frame continuum, but only the red rest-frame optical one. To do so, we considered colour selections found in the literature. In particular, we compared our sample with the LRD selection criterion by \citet{Barro2024}, which correspond to $F444W-F277W>1.5$, selecting therefore objects with a red rest-frame optical slope. As visible in Fig. \ref{fig:LRD} (top), sources in all three HELM samples occupy a completely different colour space, showing a flat or blue F277W-F444W colour, which is consistent with a simple stellar continuum at $z<1$. In the HELM-$1\sigma$ sample, there are 32 objects (2.6\% of the sample) that are too faint ($S/N<3$) in the F277W filter to efficiently derive the $F277W-F444W$ colour and could be a even fainter counterpart of the known LRDs. If we instead considered the selection criterion by \citet{Greene2024}, that is based on two colour selections (i.e., F277W-F444W and F115W-F200W), we found that no HELM sources are LRD candidates (Fig. \ref{fig:LRD}, bottom). 

Overall, we can exclude that any of the robust HELM sources are z>4 LRDs, while a minority of the HELM-$1\sigma$ sources may be, but it highly depends on the selection criteria considered. Moreover, the overall physical properties of the HELM population seem also to exclude a direct evolution of the LRDs into the HELM.

%%%%%%%%%%%%%%%%
\subsection{Are HELM galaxies more compact than non-dusty dwarfs?}
In this Section we investigate the possibility that HELM galaxies are more compact than galaxies of similar stellar mass. This would cause a large dust attenuation even if the overall dust content remains similar. To investigate this possibility, we first derived the FWHM of both HELM galaxies and non-dusty dwarfs, fitting the radial profile of each source with a Gaussian profile using the \texttt{Photutils} Python package \citep{photoutils}. We performed the analysis with the F277W filter, as non-dusty dwarfs are generally faint at longer wavelengths, while HELM galaxies are faint or undetected at shorter wavelengths. We however verified, even if with smaller number statistics, that results are consistent when using the F150W and the F444W filters. 

\begin{figure}
    \centering
    \includegraphics[width=\linewidth]{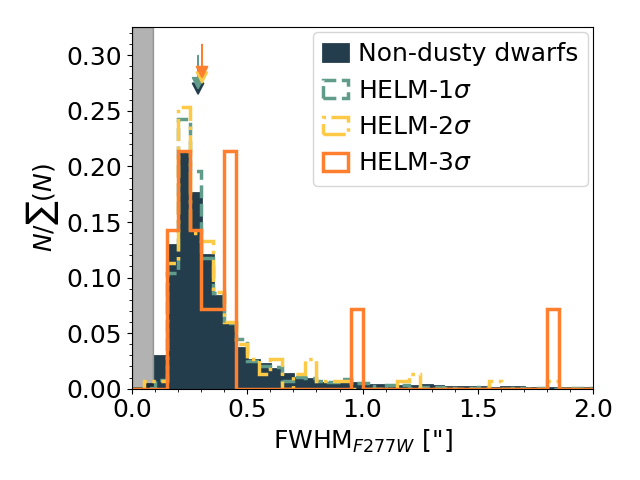}
    \caption{Comparison between the sizes of HELM galaxies and non-dusty dwarfs. We show the distributions of the FWHM in the F277W filter for the control sample of non-dusty dwarfs (dark blue filled histogram), for the HELM$-1\sigma$ sample (green dashed histogram), for the HELM$-2\sigma$ sample (yellow dash-dotted histogram), and for the HELM$-3\sigma$ sample (orange solid histogram). All samples are limited to $z<1$. Coloured arrows indicate the median of the three samples and are slightly shifted vertically for clarity. The grey shaded area shows the FWHM of the F277W filter PSF. }
    \label{fig:size}
\end{figure}

As can be seen in Fig. \ref{fig:size}, the three distributions are generally similar. The median values are 0\arcsecf28 for the non-dusty dwarfs and the HELM-$1\sigma$ sample, 0\arcsecf29 for the HELM-$2\sigma$ sample, and 0\arcsecf30 for the HELM-$3\sigma$ sample. %These values correspond to physical size between 1.3 and 1.5\,kpc, considering the median redshift of the HELM samples of $z=0.3-0.4$. 
We performed the one-sided and the two-sided KS test to exclude the possibility that the HELM samples are extracted from a distribution smaller than or coming from the same as the control sample. On the one hand, the number statistics and the difference between the HELM-$1\sigma$ sample and the sample of non-dusty dwarfs are large enough that we can exclude that the distribution of the control sample is the same or larger than the HELM-$1\sigma$ sample, with a confidence above $99\%$. This implies that $1\sigma$ HELM galaxies are not statistically smaller than the non-dusty dwarfs. On the other hand, the other two HELM samples, which suffer from low number statistics, have a p-value of 0.41 and 0.69 to exclude the possibility they are extracted from the same distribution of the non-dusty galaxies, and a p-value of 0.20 and 0.36 to exclude that the non-dusty galaxies have a greater distribution. These results indicate that we cannot exclude that HELM-$2\sigma$ and HELM-$3\sigma$ galaxies are statistically smaller than non-dusty sources.

\begin{figure}
    \centering
    \includegraphics[width=\linewidth]{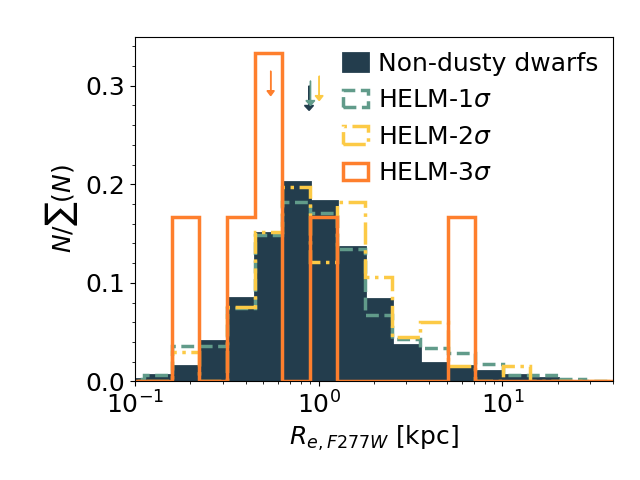}
    \caption{Comparison between the effective radius in physical scale of HELM galaxies and non-dusty dwarfs. We show the distributions in the F277W filter for the control sample of non-dusty dwarfs (dark blue filled histogram), for the HELM$-1\sigma$ sample (green dashed histogram), for the HELM$-2\sigma$ sample (yellow dash-dotted histogram), and for the HELM$-3\sigma$ sample (orange solid histogram). All samples are limited to $z<1$. Coloured arrows indicate the median of the three samples and are slightly shifted vertically for clarity. }
    \label{fig:size2}
\end{figure}

To validate further our analysis, we considered also the effective radius ($R_{e}$) estimated by McGrath et al. (in prep.) using \textsc{Galfit} \citep{Peng_2010}, converting it to physical scale using the photometric redshift.  Single-component S\'ersic profiles were fitted to all sources with $F356W < 28.5$, using empirical point-spread-functions (PSFs), generated by stacking isolated stars across all CEERS fields \citep{Finkelstein2023a}. The magnitude cut selects all sources in the HELM-$3\sigma$, 95\% of HELM-$2\sigma$ galaxies, 81\% of the HELM-$1\sigma$ sample, and 59\% of the control sample of non-dusty dwarfs, considering only sources with $z<1$. The S\'ersic profile fit was reliable (flag 0 and 1) for 6 HELM-$3\sigma$ galaxies, 66 HELM-$2\sigma$ sources, 445 HELM-$1\sigma$ objects, and 35\% of the control sample.
We show the distribution of the effective radius for the three HELM samples and the control sample in Fig. \ref{fig:size2}. The median effective radius is 0.9\,kpc for the non-dusty dwarfs and the HELM-$1\sigma$ sample, 1.0\,kpc for the HELM-$2\sigma$ sample and 0.5\,kpc the HELM-$3\sigma$ sample. We can exclude, using the one-sided KS test, that the HELM-$1\sigma$ and HELM-$2\sigma$ samples are extracted from a distribution smaller than the non-dusty dwarfs with a probability larger than 89\%, in both cases.  The statistic is too small to obtain meaningful constrains for the HELM-$3\sigma$ sample, but the hints point to these sources being indeed generally smaller. Moreover, the two-sided KS test can not discard with a high significance that all samples are extracted from the same distribution, as the comparison with the control sample gives a p-value is 0.21, 0.23 and 0.16 for the HELM-$1\sigma$, HELM-$3\sigma$, and HELM-$3\sigma$ sample, respectively.

Therefore, we can exclude that the large dust attenuation observed in HELM galaxies is driven by HELM galaxies being more compact than average for the HELM$-1\sigma$ sample and even for the HELM-$2\sigma$. The sample size is instead too small to exclude this possibility with large confidence for the HELM-$3\sigma$. These results are also limited by the spatial resolution of JWST images, as $46-50\%$ of the sources are barely resolved, having sizes below three times the FWHM of the filter PSF.

%%%%%%%%%%%%%%%%
\subsection{Are HELM galaxies preferentially edge-on?}
Another possibility is that HELM galaxies are preferentially edge-on galaxies, making the large dust attenuation a result of the line of sight. To verify this possibility we consider the axis ratio $b/a$, where $b$ and $a$ are the minor and major axis of each source, derived as part of the S\'ersic profile fit mentioned on the previous section.

\begin{figure}
    \centering
    \includegraphics[width=\linewidth]{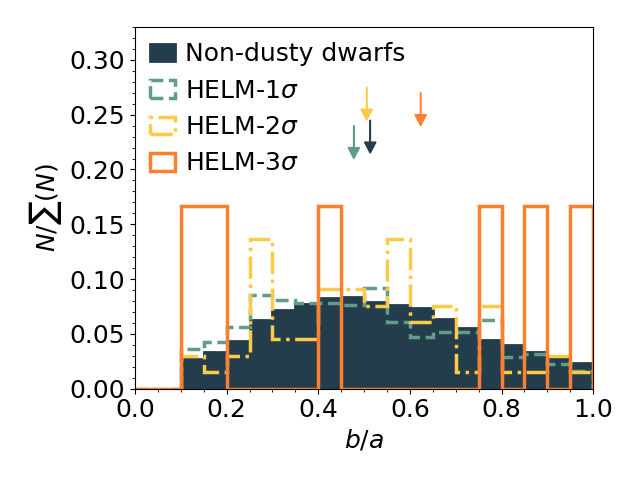}
    \caption{Comparison between the axis ratio of HELM galaxies and non-dusty dwarfs. We show the distributions of the ellipticity in the F277W filter for the control sample of non-dusty dwarfs (dark blue filled histogram), for the HELM$-1\sigma$ sample (green dashed histogram), for the HELM$-2\sigma$ sample (yellow dash-dotted histogram), and for the HELM$-3\sigma$ sample (orange solid histogram). All samples are limited to $z<1$. Coloured arrows indicate the median of the three samples and are slightly shifted vertically for clarity.}
    \label{fig:elipt}
\end{figure}

\begin{figure}
    \centering
    \includegraphics[width=\linewidth]{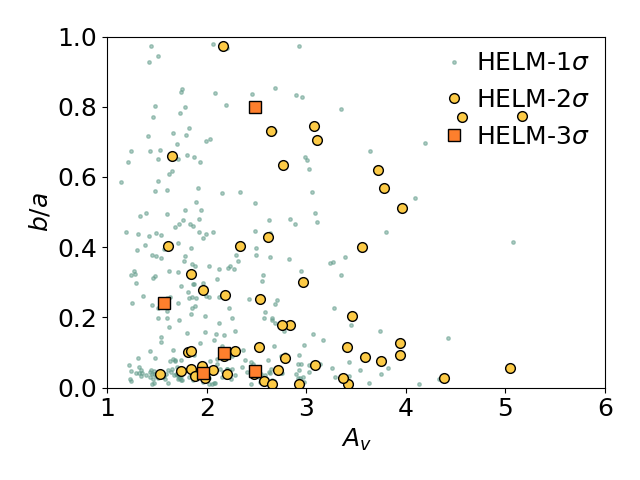}
    \caption{Dust attenuation vs. ellipticity of the HELM galaxies, as derived considering the F277W filter. There is no evidence for a correlation between the dust attenuation and the ellipticity. We limit each sample to $z<1$.}
    \label{fig:Ave}
\end{figure}

Figure \ref{fig:elipt} shows the distribution of the axis ratio for the HELM samples as well as the control sample. As for the sizes, we show the results using the F277W filter, as dusty dwarf are generally undetected at shorter wavelengths. The median axis ratio of the control sample is $b/a=0.51$, with the HELM samples having similar median values, that is $b/a=0.47$ for the HELM-$1\sigma$, $b/a=0.51$ for the HELM-$2\sigma$, and $b/a=0.62$ for the HELM-$3\sigma$. We verified the statistical significance of these differences using the one-sided and the two-sided KS tests, but we can only exclude that the HELM-$1\sigma$ sample is extracted from a greater or similar distribution than the control sample with a confidence of 98\%. The absence of a correlation between the dust attenuation and the axis ratio (Fig. \ref{fig:Ave}) supports that the HELM galaxies are not preferentially edge-on galaxies.

\citet{Pandya202} have shown that disk-on oblate systems preferentially have larger dust attenuation. We can exclude these possibility only for 30--40\% of the HELM samples, as the derived axis ratio is larger than the one expected for disk-on oblate systems ($b/a>0.4$). However, the same fraction of large $b/a$ ratio is also present among the non dusty dwarfs.

Another possibility is that these galaxies are prolate systems, with the major axis pointing towards us. Studies have shown that dry major mergers could originate prolate galaxies, but their fraction seems to also strongly decrease with decreasing stellar mass \citep[e.g.][]{Tsatsi2017,Li2018}. There are some rare case of dwarf spheroidal galaxies showing prolate shapes in the local Universe \citep[e.g.][]{Ebrova2015}, while their fraction seems to increase with redshift. However, we can not test these scenario for HELM galaxies using only photometric data, due to the degeneracies to convert projected shapes into intrinsic ones and their faintness. 

We can therefore conclude that with the current data seems that an inclination effect is probably not the primary cause of the large dust attenuation observed in HELM galaxies, even though observations with higher angular resolution are necessary to have more robust conclusions and spatially resolved spectroscopic data are necessary to investigate possible prolate shapes.

%%%%%%%%%%%
\subsection{Are HELM galaxies linked to galaxy interactions?}

A galaxy interaction could potentially increase the SFR of a galaxy, boosting also metal and dust production. Another possibility is that HELM galaxies are tidal dwarfs \citep[e.g.][]{Zwicky1956,Duc1999}, forming from debris in interacting or merging galaxies. While these scenarios differ in many aspects, such as the absence of non-baryonic dark matter in tidal dwarfs \citep{Lelli2015}, they both require HELM galaxies to be in over-dense regions. A good estimation of the galaxy environment of HELM galaxies would require measuring their spectroscopic redshift and the one of their neighbours over a large area. At the moment this is not available, so we estimate the projected distance of the $5^{th}$ neighbour ($d_{5th}$) as a proxy for galaxy environment. For this estimation, we consider all reliable galaxies in the catalogue with photometric redshift consistent with the one of HELM galaxies, withing the $1\sigma$ uncertainties. \\ 

\begin{figure}
    \centering
    \includegraphics[width=\linewidth]{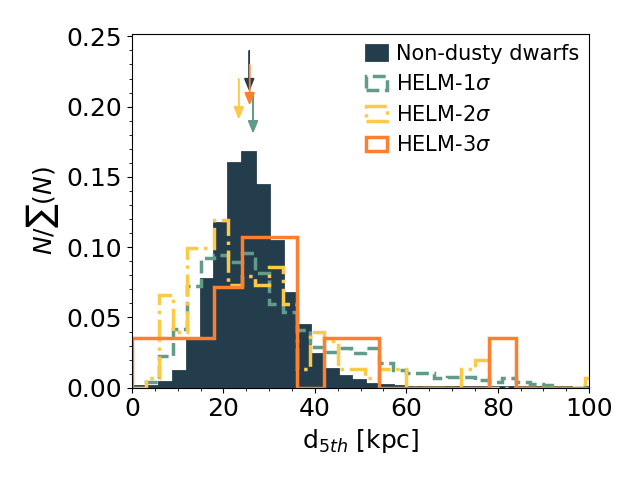}
    \caption{Distance of the $5^{th}$ neighbour, used as proxy for galaxy clustering. We considered all sources with photometric redshift consistent with the photometric redshift of HELM galaxies, within 1$\sigma$ uncertainty. We consider the photometric redshift of HELM sources to convert from angular to physical scale. Histograms are shown for the control sample of non-dusty dwarfs (dark blue filled histogram), for the HELM$-1\sigma$ sample (green dashed histogram), for the HELM$-2\sigma$ sample (yellow dash-dotted histogram), and for the HELM$-3\sigma$ sample (orange solid histogram). All samples are limited to $z<1$. Coloured arrows indicate the median of the three samples and are slightly shifted vertically for clarity.}
    \label{fig:clustering}
\end{figure}

As shown in Figure \ref{fig:clustering}, there is little difference in the median of the four distribution, that is $d_{5}=25.7\,\rm kpc$, 27.6\,kpc, 22.6\,kpc, and 25.7\,kpc, for the sample of non-dusty dwarfs, HELM$-1\sigma$, HELM$-2\sigma$, and HELM$-3\sigma$, respectively. However, the distributions of HELM$-1\sigma$ and HELM$-2\sigma$ sources are slightly more skewed to lower $d_{5}$. Indeed, the one-sided and two-sided KS tests discarded with a confidence larger than 99.9\% that the distribution of not-dusty dwarfs is smaller or the same as the ones of HELM$-1\sigma$ and HELM$-2\sigma$ samples. Again, maybe because of the limited statistic, we can not draw similar conclusions for the HELM$-3\sigma$ sample.  

Given the dependence of clustering on stellar mass, we more carefully verify that this discrepancy is not driven by variations in the stellar mass or redshift distributions of the non-dusty and HELM samples. For this reason, for each source in the HELM samples we selected three non-dusty dwarfs having the minimum difference in redshift and stellar mass. The median distance of the $5^{th}$ neighbour of these non-dusty dwarfs is similar to the previous ones, as they become $d_{5}=26.5\,$kpc, $d_{5}=23.3\,$kpc, and $d_{5}=25.7\,$kpc for the subsamples matched to the HELM$-1\sigma$, HELM$-2\sigma$, and HELM$-3\sigma$ samples, respectively. 

\begin{figure}
    \centering
    \includegraphics[width=\linewidth]{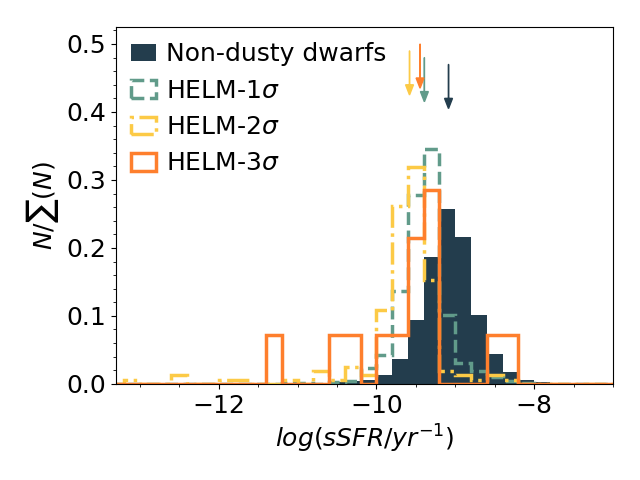}
    \caption{HELM galaxies are generally less star-forming than their non-dusty counterparts. We show the sSFR histograms of the control sample of non-dusty dwarfs (dark blue filled histogram), of the HELM$-1\sigma$ sample (green dashed histogram), of the HELM$-2\sigma$ sample (yellow dash-dotted histogram) and of the HELM$-3\sigma$ sample (orange solid histogram). Coloured arrows indicate the median of the three samples and are slightly shifted vertically for clarity. All samples are limited to $z<1$.}
    \label{fig:ssfr}
\end{figure}

Based on the results on the HELM$-1\sigma$ and HELM$-2\sigma$ samples, some HELM galaxies could be more clustered than their non-dusty counterparts. In this case, we need to understand if they are going through an interaction or if they are the product of interactions, like tidal dwarfs \citep[e.g.][]{Zwicky1956,Duc1999}. Dynamical studies to verify the presence or absence of a dark matter halo and spectroscopic studies to derive gas abundances are beyond the possibilities of this work and JWST angular resolution limits any detailed morphological study. However, we can analyse if these galaxies are going through a burst of star-formation, possibly induced by galaxy interactions. \citet{Bisigello2023b} have already shown, although using a selection based on colours and not on physical properties, that this scenario seems to be excluded. We confirm their finding in Fig. \ref{fig:ssfr}. Indeed, the median sSFR of the HELM samples is smaller than the value of the non-dusty dwarfs, being $\rm log_{10}(sSFR/yr^{-1})=-9.1$ for the non-dusty dwarfs and from $\rm log_{10}(sSFR/yr^{-1})=-9.6$ to $\rm log_{10}(sSFR/yr^{-1})=-9.4$ for the HELM samples. The two-sided KS test shows that we can exclude with a confidence larger than 99.9\% that the HELM samples are derived from the same distribution as the non-dusty dwarfs. Therefore, we can exclude that the large dust attenuation of HELM galaxies is driven by a burst of star-formation. It is however necessary to take into account that the sSFR estimation is based on SED fitting with limited photometry, given that both the non-dusty dwarfs and the HELM galaxies are generally faint. In addition, considering the available data, the possibility that these objects are instead post-starburst galaxies is still totally open.

%%%%%
\section{Conclusions}
Using CEERS data, we selected three classes of HELM galaxies, considering the fraction of their multiparameter distributions that fall within the HELM selection, as derived through SED fitting and using a variety of configurations. We then cross-matched these samples with catalogues of $z>8.5$ galaxies and brown-dwarf, to remove possible contaminants. We also removed any source for which the SED fitting preferred a $A_V<1$~mag solution and we visually inspected all remaining sources. We obtained 14 very robust HELM candidates (HELM-$3\sigma$), with a probability $>99.7\%$ of being HELM sources, 158 robust candidates (HELM-$2\sigma$), with a probability between 95 and 99.7\%, and 1189 plausible HELM galaxies (HELM-$1\sigma$), with a probability between 68 and 95\%.

The large majority (90\%) of the HELM sources are at $z<1$, indicating they are mainly a low-$z$ population. They have a median stellar mass of $10^7\,\rm M_{\odot}$ and a median dust attenuation of $A_V=2$~mag. However, the only galaxies confirmed spectroscopically are at $z>4$. This may be due to selection biases, as these were previously identifies as $z>10$ candidates, or indicate a general failure in the SED fitting analysis to systematically derive their redshifts. Spectroscopic follow-ups are therefore pivotal to clarify the redshifts of these galaxies.

The stacked photometry of HELM sources, compared with a control sample of non-dusty dwarfs, reassured us of the red nature of these galaxies, showing also a tentative $3.3\,\mu m$ PAH feature. At the same time, the use of various colour selections can exclude that the majority of these sources are bright little red dots, while we cannot exclude that they are faint ones. We can also exclude that the majority of these sources are strong candidates at $z>15$.

Examining the morphology of HELM sources, while bearing in mind that they are faint and relatively compact, we can rule out that the inclination has a major factor contributing to the high dust attenuation observed in HELM galaxies.
Moreover, the distributions of sizes of the HELM subsamples and the control sample do not show any statistical difference. At the same time, with the current data we can not investigate if HELM sources are prolate systems.

The analysis of the environment shows that the distribution of HELM galaxies (HELM$-2\sigma$ and HELM$-1\sigma$) may be slightly skewed to more dense environment, having however similar median than the control sample.
While spectroscopic follow-ups are fundamental to confirm this statement, this may indicate that HELM galaxies are the result of galaxy interactions (e.g., tidal dwarfs, stripped galaxies, merger-triggered starbursts). However, there seems to be no indication of a burst of star formation, as HELM sources are on average less star-forming than non-dusty dwarfs, but there is still the possibility that they are post-starburst galaxies. 

Considering that HELM galaxies represent only 1.3\% of the entire CEERS sample and maximum 5\% of the galaxies with similar redshift and stellar mass, they could also be a short-lived evolutionary phase: AGB stars may have saturated these galaxies with carbon-based dust, while not enough supernovae have gone off to blow all the dust out, or at least to create enough holes in it.

In the future, to have a more in-depth analysis of the morphology of HELM galaxies to unveil any possible signature of galaxy interactions we need to wait for the upcoming Extremely Large Telescope (ELT), having a PSF $\approx 6$ times sharper than JWST. At the same time, to verify that dust attenuation is not purely a geometrical effect, it will be necessary to observe these galaxies in the far-IR to trace their dust continuum. HELM sources are expected to be too faint to be observable with the Northern Extended Millimetre Array (NOEMA), having an average flux of 0.55\mum at 1.1\,mm and 0.03\mum at 3\,mm. At least for the brightest HELM candidates, follow-ups should instead be possible with the future NASA Probe far-IR Mission for Astrophysics (PRIMA, P.I. J. Glenn), as shown by \citet{Bisigello2025b}, or with the Atacama Large Millimeter Array (ALMA) for HELM sources in southern fields. Finally, a more in depth study of these intriguing population is pivotal to also have better control of the main contaminants of ultra high-$z$ galaxies \citep{Gandolfi2025,Castellano2025}.

\begin{acknowledgements}
The research activities described in this paper were carried out with contribution of the Next Generation EU funds within the National Recovery and Resilience Plan (PNRR), Mission 4 -- Education and Research, Component 2 -- From Research to Business (M4C2), Investment Line 3.1 -- Strengthening and creation of Research Infrastructures, Project IR0000034--“STILES -- Strengthening the Italian Leadership in ELT and SKA”. GR, GG and AG are supported by the European Union – NextGenerationEU RFF M4C2 1.1 PRIN 2022 project 2022ZSL4BL
INSIGHT. GR and LB acknowledge the support from MIUR gran PRIN 2017 20173ML3WW-0013. GR, LB and MG acknowledge support from INAF under the Large Grant 2022 funding scheme (project “MeerKAT and LOFAR Team up: a Unique Radio Window on Galaxy/AGN co-Evolution”). MG acknowledges support from INAF under the the following funding scheme Large GO 2024 (project "MeerKAT and Euclid Team up: Exploring the galaxy-halo connection at cosmic noon")". LB also acknowledges support from the INAF Large Grant 2022 “Extragalactic Surveys with JWST” (PI Pentericci).
This research made use of Photutils, an Astropy package for detection and photometry of astronomical sources \citep{photoutils}.
\end{acknowledgements}

\bibliographystyle{aa} % style aa.bst
\bibliography{main} % your references Yourfile.bib
\begin{appendix}

\section{Comparison between different SED fitting runs} \label{sec:sedcomparison}
\begin{figure}
    \centering
    \includegraphics[width=\linewidth,trim={23 20 20 20},clip]{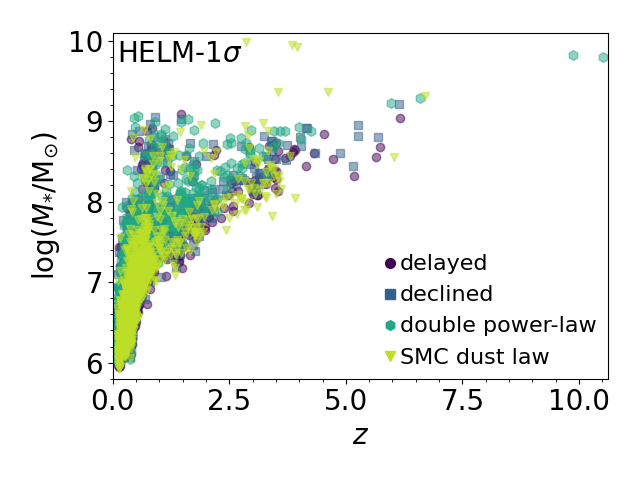}
    \includegraphics[width=\linewidth,trim={10 20 12 20},clip]{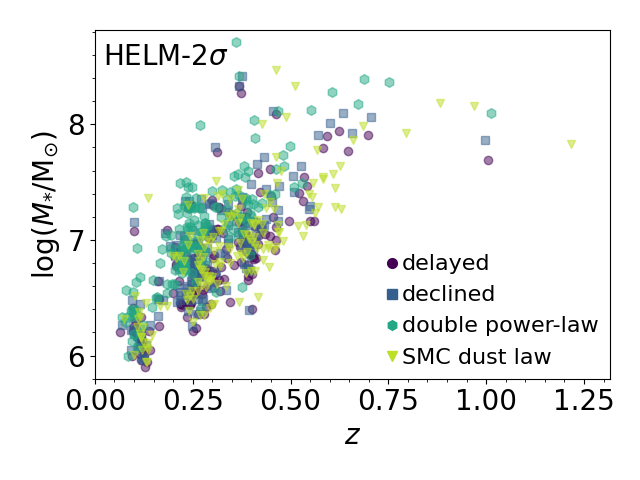}
    \includegraphics[width=\linewidth,trim={10 20 12 20},clip]{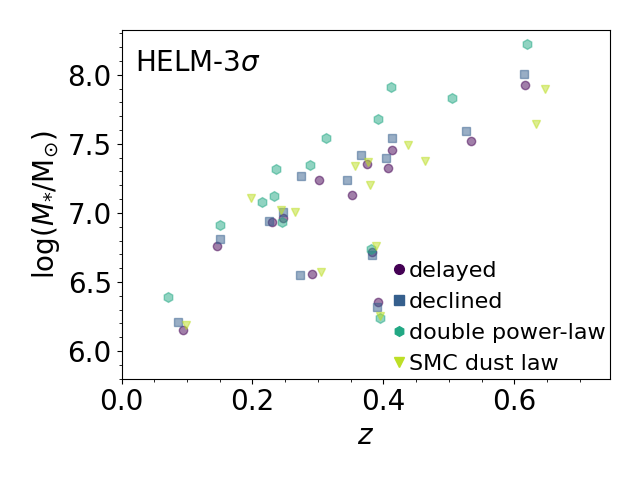}
    \caption{Comparison between the redshift and the stellar mass obtained with the different SED runs. The comparison is performed for the HELM-$1\sigma$ (top), the HELM-$2\sigma$ (centre), and the HELM-$3\sigma$ samples. Different colours indicate different SED fitting configurations: delayed SFH (pink circles), declined SFH (grey squares), double power-law SFH (olive hexagons), and delayed SFH with SMC dust attenuation law (cyan triangles). For the HELM-$2\sigma$ we did not show, for clarity, three data points at $z=4.5$ and $z=10.6$, obtained with the delayed SFH with both reddening laws and the double power-law SFH, respectively.}
    \label{fig:zM}
\end{figure}

In this Appendix we report on the comparison between the stellar mass and redshift derived with the different SED fitting configurations (see Sec. \ref{sec:SEDfitting}) for the three HELM samples. The comparison is shown in Fig. \ref{fig:zM}. \par

On average, the delayed SFH produces the lowest stellar mass, while the double power-law SFH produces the highest, but the average difference remains below 0.09 dex. On the contrary, the double power-law SFH corresponds to the smallest redshift, while the delayed SFH with the SMC reddening laws results in the highest one. Also in this case the differences are negligible (i.e., $\Delta z<0.09$). \par

Is is worth noticing that in the case of the HELM-$1\sigma$ sample there are some solutions with $z>7$. This happens for 2 objects with the double power-law SFH.

\section{Visual inspection}\label{sec:vis}

\begin{figure}[h!]
    \centering
    \includegraphics[width=\linewidth,trim={70 20 70 0},clip]{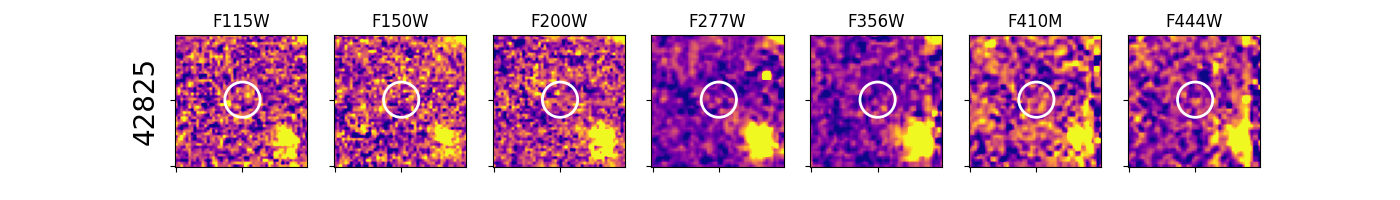}
    \includegraphics[width=\linewidth,trim={70 20 70 20},clip]{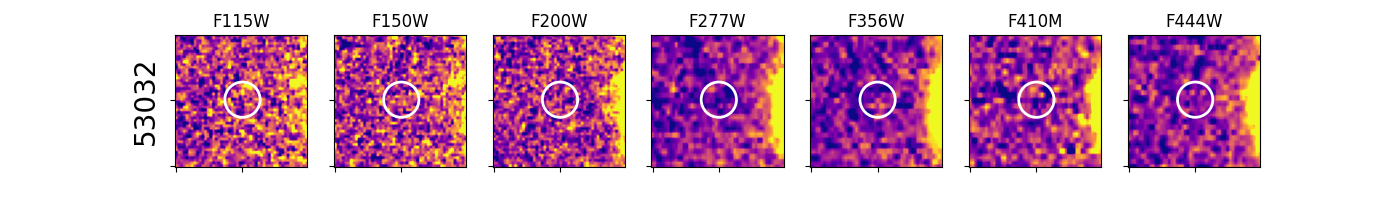}
    \includegraphics[width=\linewidth,trim={70 20 70 20},clip]{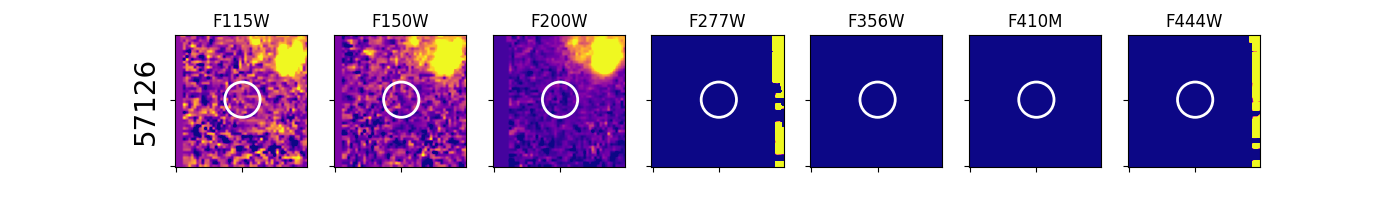}
    \caption{NIRCam cutouts of the three HELM$-3\sigma$ sources removed after visual inspection. Each cutout is $1\arcsecf5\times1\arcsecf5$ and the white circle indicates the position of the source.}
    \label{fig:vis}
\end{figure}

\begin{figure}[h!]
    \centering
    \includegraphics[width=\linewidth,trim={70 20 70 0},clip]{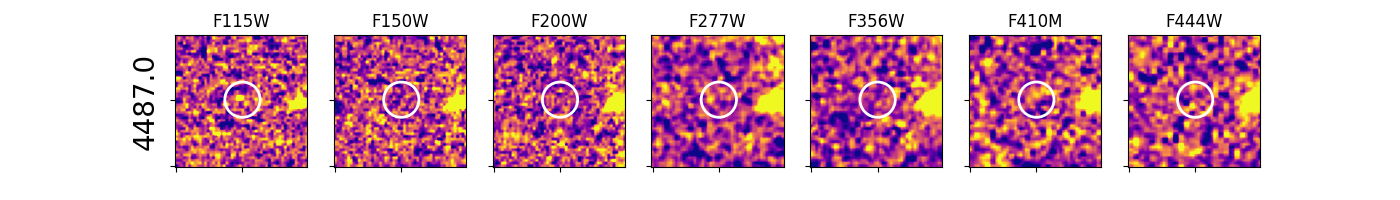}
    \includegraphics[width=\linewidth,trim={70 20 70 20},clip]{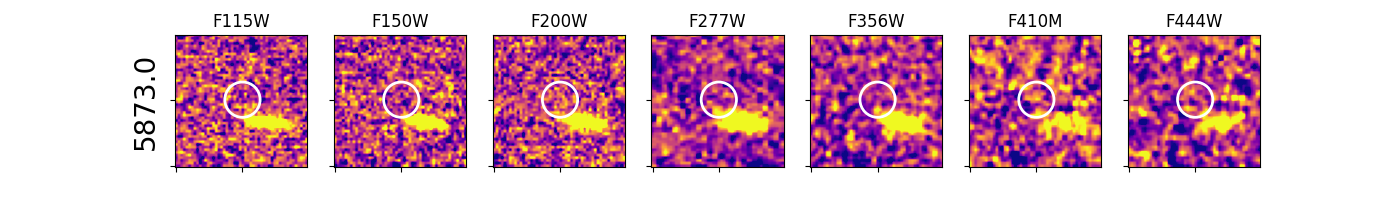}
    \includegraphics[width=\linewidth,trim={70 20 70 20},clip]{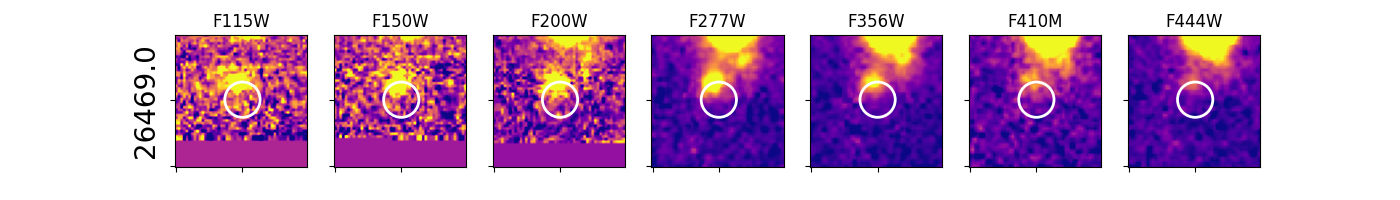}
    \includegraphics[width=\linewidth,trim={70 20 70 20},clip]{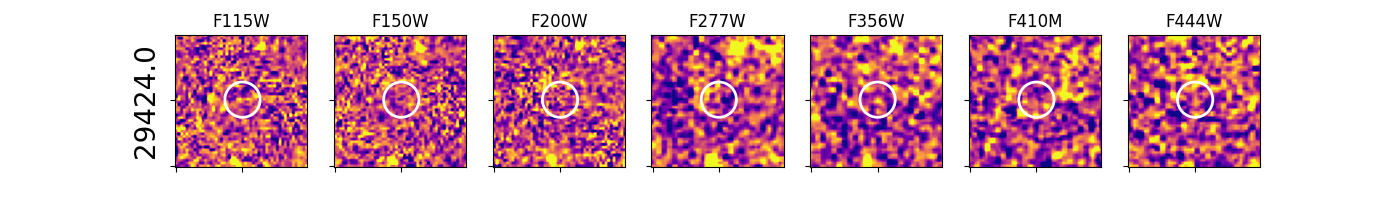}
    \includegraphics[width=\linewidth,trim={70 20 70 20},clip]{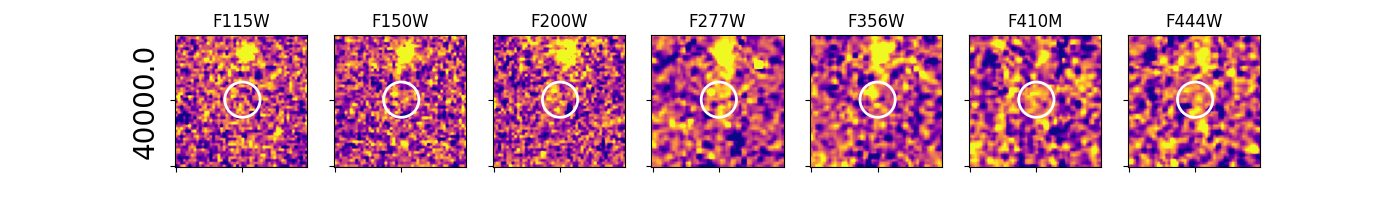}
    \includegraphics[width=\linewidth,trim={70 20 70 20},clip]{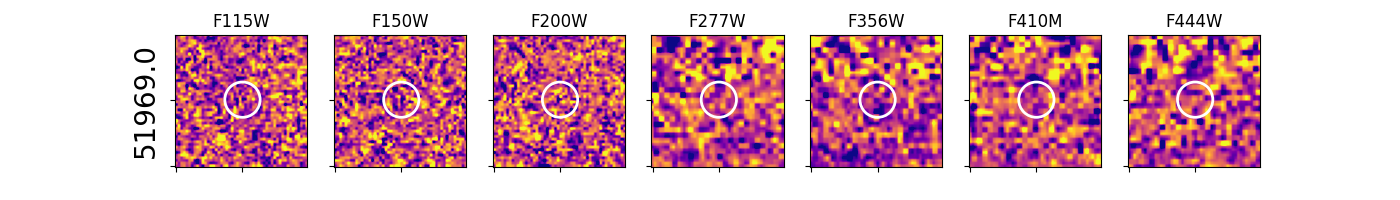}
    \includegraphics[width=\linewidth,trim={70 20 70 20},clip]{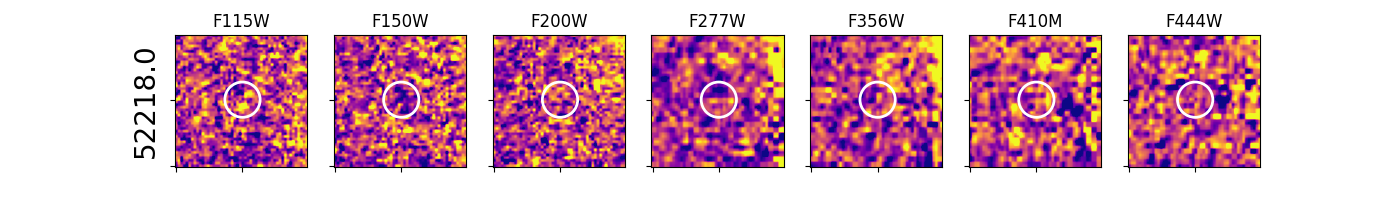}
    \includegraphics[width=\linewidth,trim={70 20 70 20},clip]{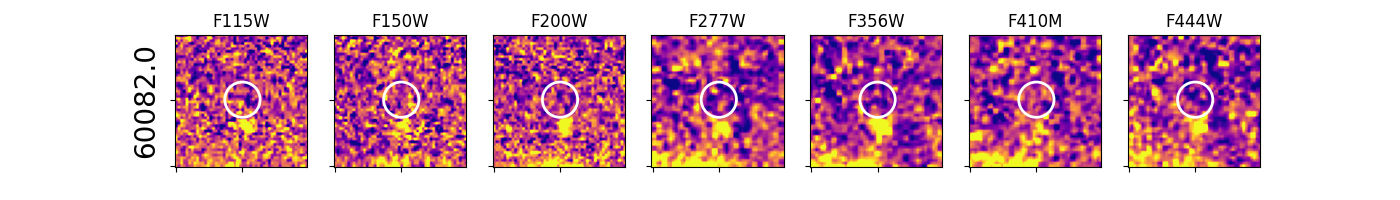}
    \includegraphics[width=\linewidth,trim={70 20 70 20},clip]{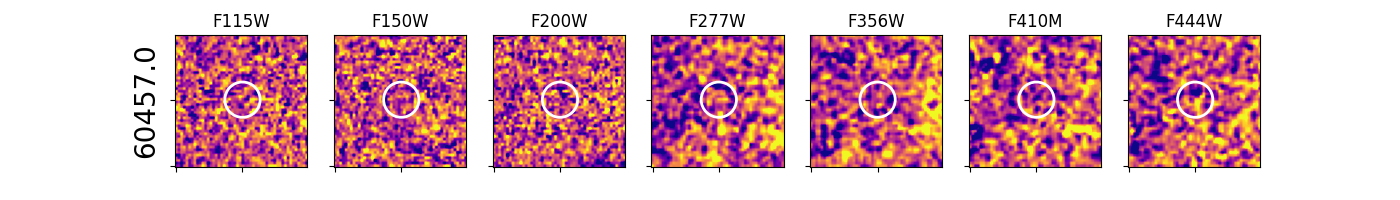}
    \includegraphics[width=\linewidth,trim={70 20 70 20},clip]{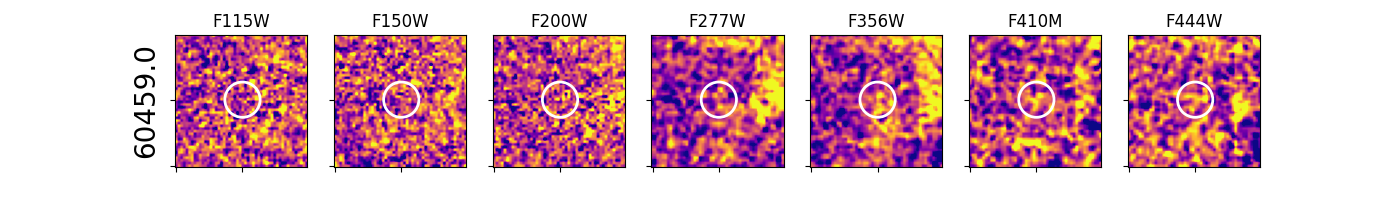}
    \includegraphics[width=\linewidth,trim={70 20 70 20},clip]{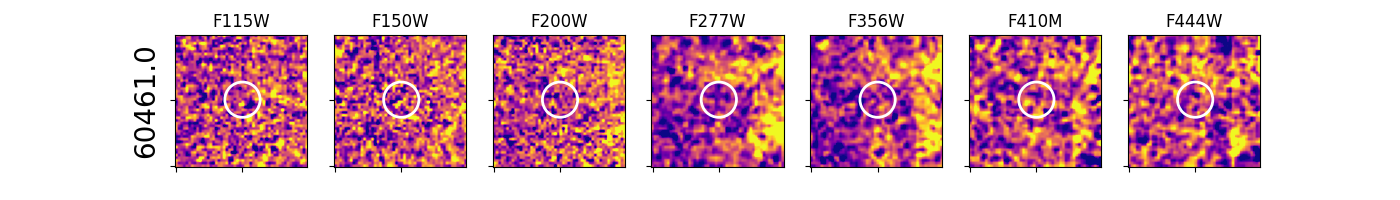}
    \includegraphics[width=\linewidth,trim={70 20 70 20},clip]{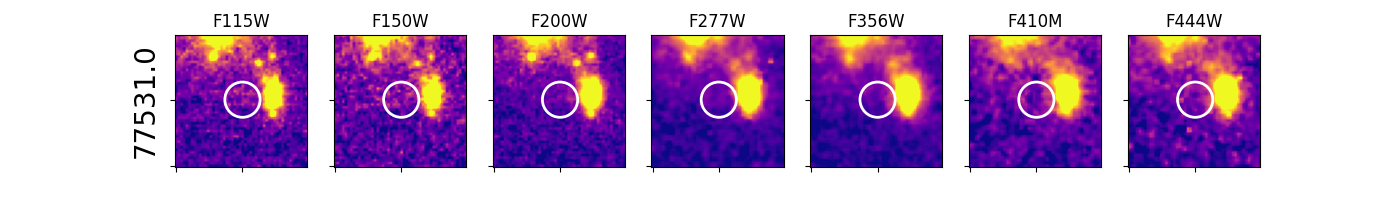}
    \includegraphics[width=\linewidth,trim={70 20 70 20},clip]{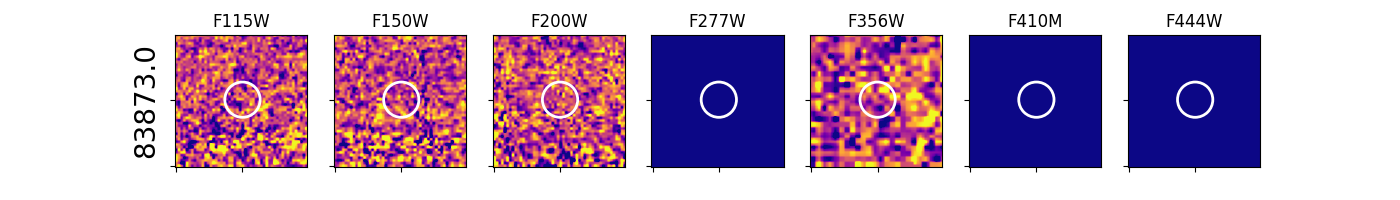}
    \includegraphics[width=\linewidth,trim={70 20 70 20},clip]{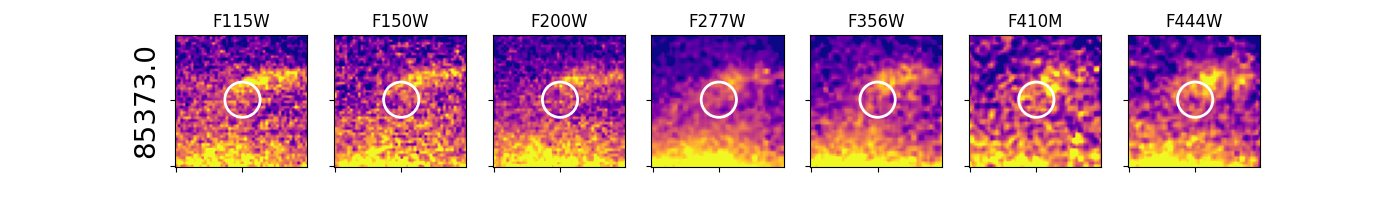}
    \includegraphics[width=\linewidth,trim={70 20 70 20},clip]{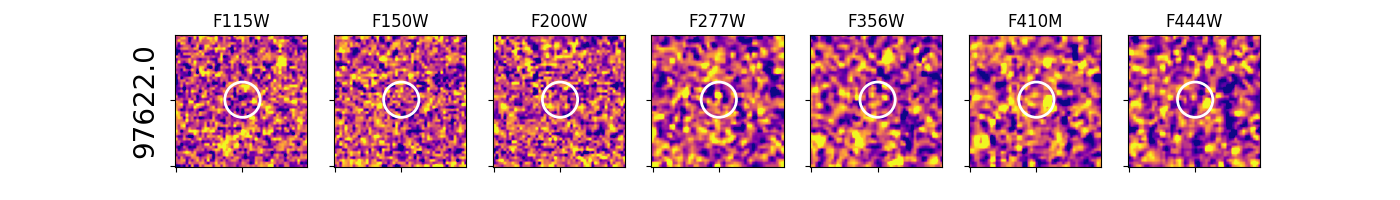}
    \caption{ NIRCam cutouts of the fifteen HELM$-2\sigma$ sources removed after visual inspection. Each cutout is $1\arcsecf5\times1\arcsecf5$ and the white circle indicates the position of the source.}
    \label{fig:vis2}
\end{figure}

In Figures \ref{fig:vis} and \ref{fig:vis2} we reported, as example, the cutouts of the sources in the HELM$-3\sigma$ and HELM$-2\sigma$ that we removed after visual inspection. We do not report the cutouts for the HELM$-1\sigma$ given the size of the sample. 
As visible in the Figures, there are different reasons that motivate our choice to remove these sources. In some cases no source is visible at the expected position, while in other cases the source has a strange shape which sometimes varies between filters. In other, more simple cases, the source is simply outside the field of view of some NIRCam filters.

\section{Available spectroscopic data}\label{sec:spec}
As mentioned in Section \ref{sec:selection}, we matched our photometrically selected samples with the available spectroscopic data to remove possible contaminants or validate HELM sources. This step was performed before creating the final HELM samples and we discussed in this Section the analysed sources. Spectroscopic data is
available as part of the CEERS survey and from the DAWN JWST Archive. The first spectroscopic data consist of slitless NIRCam WFSS spectra \citep[N. Pirzkal, private communication; see ][for grism data reduction details]{Pirzkal2018} and NIRSPEC MSA PRISM and medium resolution spectra taken as part of programme DD-2750 \citep[P.I. Arrabal Haro][]{ArrabalHaro2023b}. Given the faintness of our sample, we do not expect many galaxies to be present in the NIRCam WFSS sample, which includes only galaxies with $\rm F356W\leq25$ mag. 
Indeed, only three HELM candidates, which are in the HELM-$1\sigma$ sample, have slitless spectra, but the signal-to-noise ratio is so low that neither spectral lines nor continuum features are detectable.

There are instead ten objects that have NIRSpec PRISM data, one in the HELM-$3\sigma$ sample, one in the HELM-$2\sigma$ sample, and eight in the HELM-$1\sigma$. From the DJA we retrieved other eight NIRSpec spectra, seven from the Red Unknowns: Bright Infrared Extragalactic Survey \citep[RUBIES;][]{deGraaff2024} and one from the Cycle 1 GO-2565 (P.I. K. Glazebrook), which however is not among the massive quiescent galaxies described by \citet{Nanayakkara2025}. Of these sources, seven are in the HELM-$1\sigma$ sample and one is in the HELM-$2\sigma$ one. Overall, we have spectroscopic data for one HELM-$3\sigma$ source that was rejected, two HELM-$2\sigma$ sources, one of which was confirmed while the other was rejected, and eleven HELM-$1\sigma$ sources, nine of which were rejected, while the data were not discriminating enough for two galaxies. We discuss them in more detail in the rest of this section.

\subsection{HELM-$3\sigma$ spectrum}
\begin{figure}
    \centering
    \includegraphics[width=\linewidth]{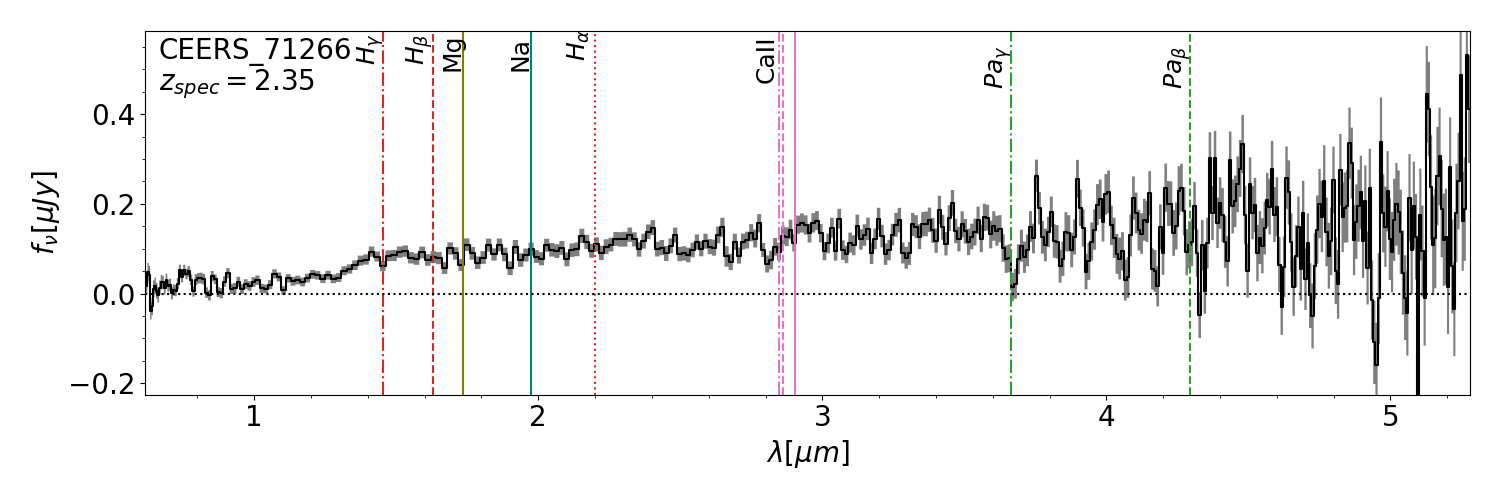}
    \caption{The only source in the HELM-$3\sigma$ sample with NIRSpec spectroscopic data. No emission lines are clearly visible, while there are some possible absorption features.}
    \label{fig:s3_spectra}
\end{figure}

% \subsubsection*{CEERS-71266}
Figure \ref{fig:s3_spectra} shows the only object in the HELM-$3\sigma$ sample with spectroscopic data, which is CEERS-71266. As visible, there are no evident line emissions, suggesting moderate star formation, but the stellar continuum is visible at $\lambda>1.4\,\rm\mu m$. There are some faint absorption features, which combined with the visible continuum break suggest that the galaxy may be located at $z=2.35$. While more data are necessary to more robustly validate this possibility, when performing a photometric fit with \textsc{Bagpipes}, using a setup similar to the one in Section \ref{sec:SEDfitting} with a delayed SFHs and fixing the redshift to $z=2.35$, we obtained a stellar mass of ${\rm log_{10}}(M_{*}/{\rm M_{\odot}})=9.3\pm0.1$ and a dust attenuation of $A_V=0.4\pm0.2$~mag. Given that this galaxy is outside the HELM selection we removed it from the final HELM sample.

\subsection{HELM-$2\sigma$ spectra}
There are two sources with spectroscopic data that were originally in the HELM-$2\sigma$ sample, one of them has been confirmed, while the other is a contaminant.

\subsubsection*{CEERS-14821}
The first source we discuss in the -$2\sigma$ sample is CEERS-14821, which has been observed as part of the Director's Discretionary Time programme JWST-DDT 2750 \citep[PI Arrabal Haro;][]{ArrabalHaro2023b} and has been presented in detail in \citet{Bisigello2025a}. The object has a spectroscopic redshift of $z_{spec}=4.883\pm0.003$ based on the $[\ion{O}{III}]5007$\AA\xspace line, which is the brightest. The $H_\alpha$, $H_\beta$, $H_\gamma$ and $[\ion{O}{III}]4959$\AA\xspace nebular emission lines are also detected. A careful spectro-photometric fit of the available data is presented in \citet{Bisigello2025a} and confirms the HELM nature of this source, which has $A_V=2.2_{-0.6}^{+0.5}$~mag, considering the reddening law by \citet{Calzetti2000}, and a stellar mass of ${\rm log_{10}}(M_*/\rm M_{\odot})=8.17_{-0.04}^{+0.05}$. 

CEERS-14821 is at very high-$z$ with respect to the rest of the HELM population, so it may be not representative of the entire population. It however advocates on the necessity of spectroscopic follow-ups, to verify the nature and properties of these intriguing sources.

\subsubsection*{CEERS-46775}
\begin{figure}
    \centering
    \includegraphics[width=\linewidth]{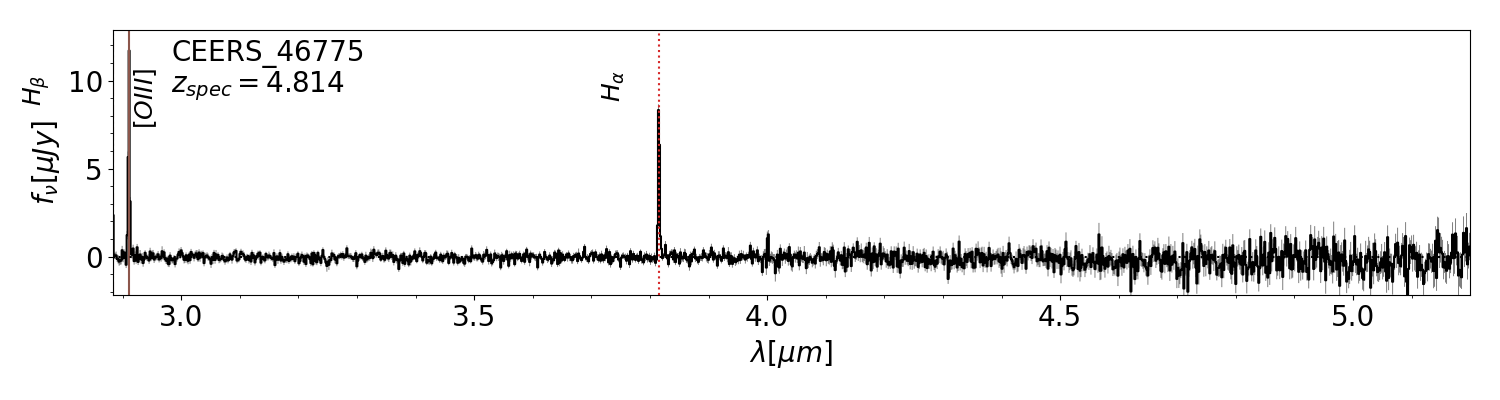}
    \caption{CEERS-45766, a source in the HELM-$2\sigma$ sample with medium resolution (G395M) NIRSpec data. The other source in the HELM-$2\sigma$ catalog was presented in \citet{Bisigello2023b}.}
    \label{fig:s2_spectra}
\end{figure}

The second galaxy in the HELM-$2\sigma$ sample with spectroscopic data is CEERS-46775. This galaxy has been observed as part of the RUBIES survey with the G395M/F290LP grating and filter. This source has a spectroscopic redshift of $z=4.81353\pm0.00008$ derived from the [\ion{O}{III}]5007\AA\xspace and $H_{\alpha}$ emission lines (Fig. \ref{fig:s2_spectra}). The [\ion{O}{III}]4959\AA\xspace is at the edge of the spectrum, while $H_{\beta}$ is outside the wavelength coverage, preventing us from measuring the dust extinction directly from the Balmer decrement.
We performed a spectro-photometric fit with bagpipes, fixing the redshift to the spectroscopic one and considering a non parametric SFH. We this setup, we derived a stellar mass of ${\rm log_{10}}(M_*/\rm M_{\odot})=9.13_{-0.04}^{+0.25}$ and a dust attenuation of $A_V=0.03_{-0.2}^{+0.05}$~mag, completely outside the HELM selection. We therefore remove this object from the HELM-$2\sigma$ sample.

\subsection{HELM-$1\sigma$ spectra}
Eleven sources originally present in the HELM-$1\sigma$ have NIRSpec spectroscopic data. None of them were confirmed as HELM sources.

\begin{figure}
    \centering
    \includegraphics[width=\linewidth,trim={5 50 0 0},clip]{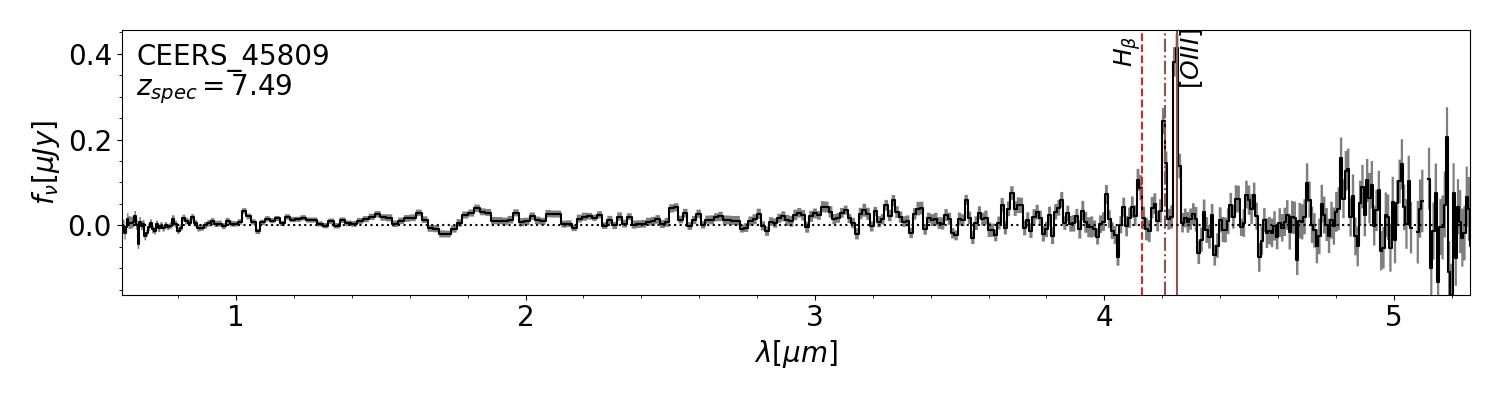}
    \includegraphics[width=\linewidth,trim={20 50 0 0},clip]{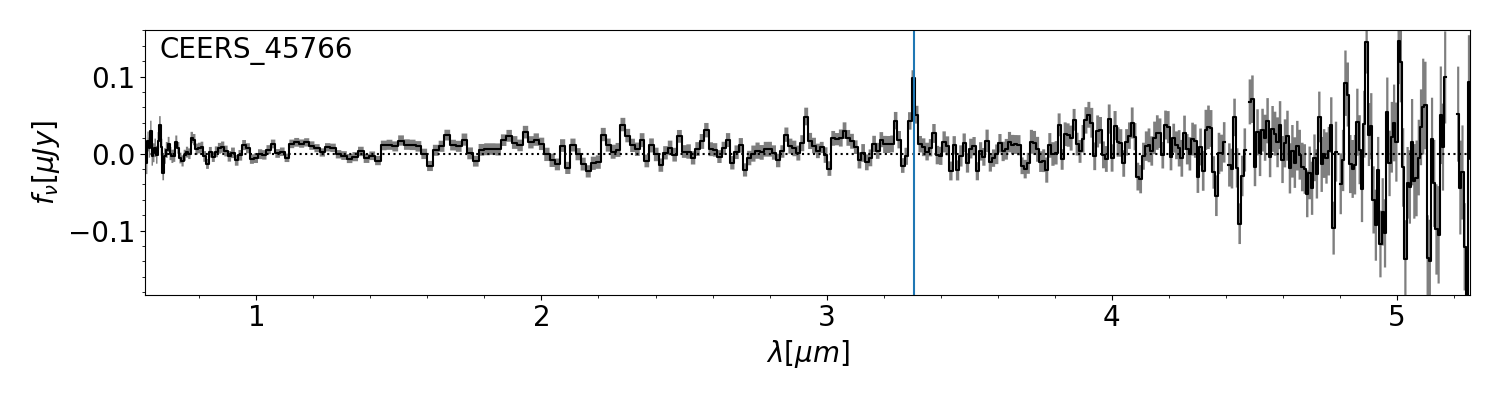}
    \includegraphics[width=\linewidth,trim={20 50 0 0},clip]{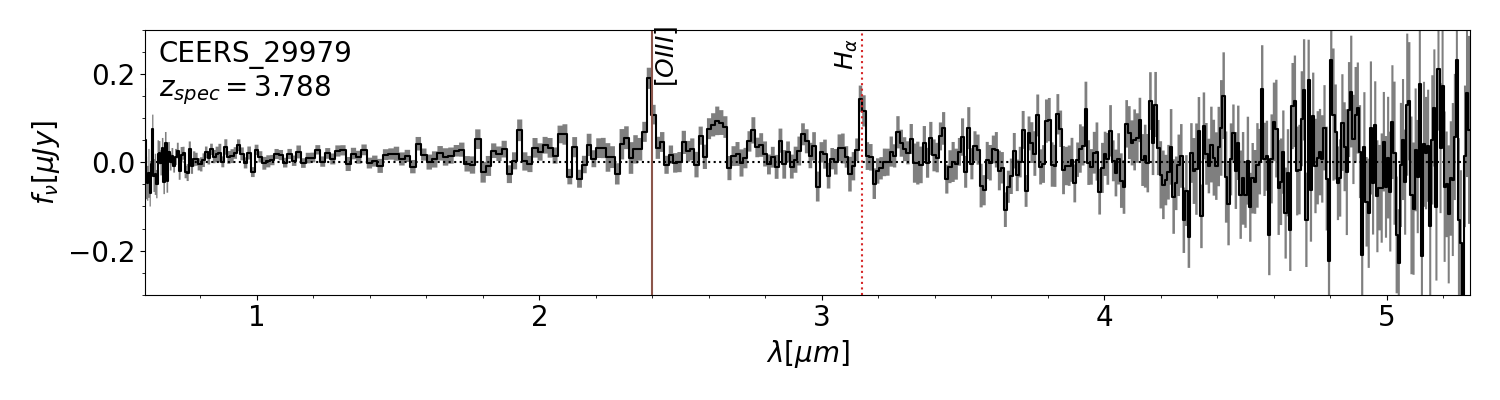}
    \includegraphics[width=\linewidth,trim={20 50 0 0},clip]{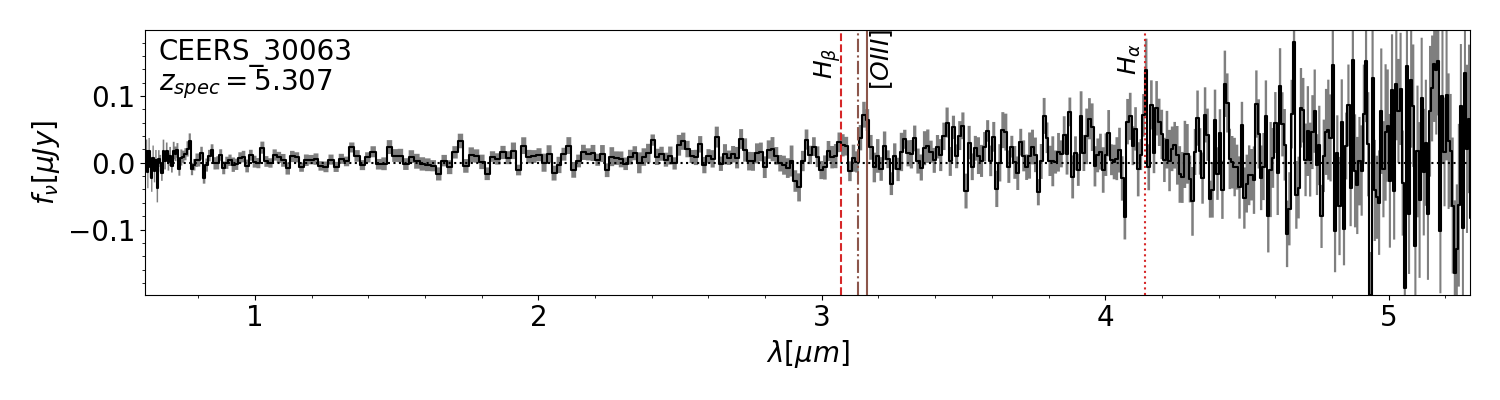}
    \includegraphics[width=\linewidth,trim={5 50 0 0},clip]{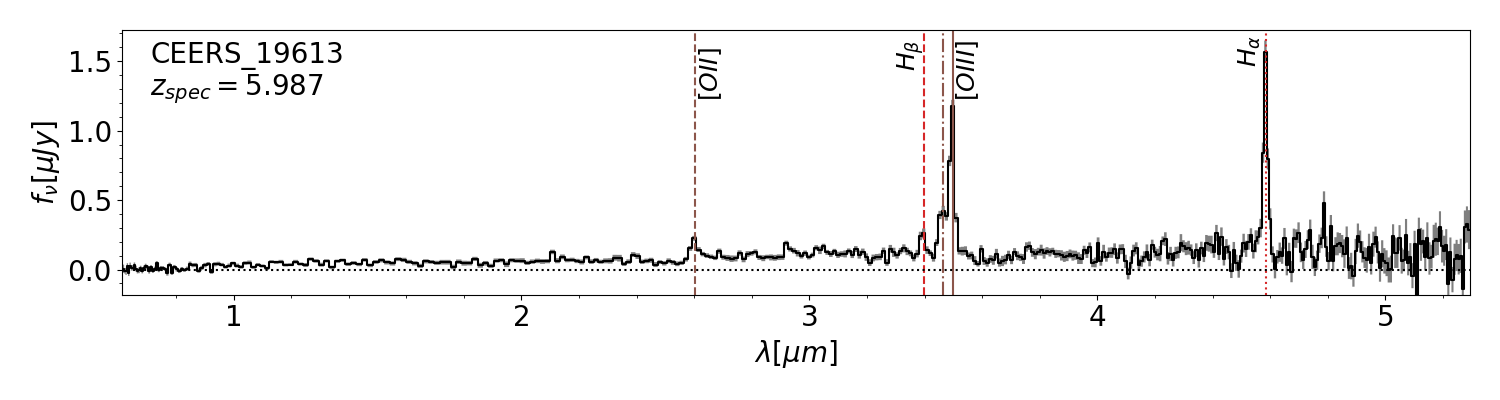}
    \includegraphics[width=\linewidth,trim={5 50 0 0},clip]{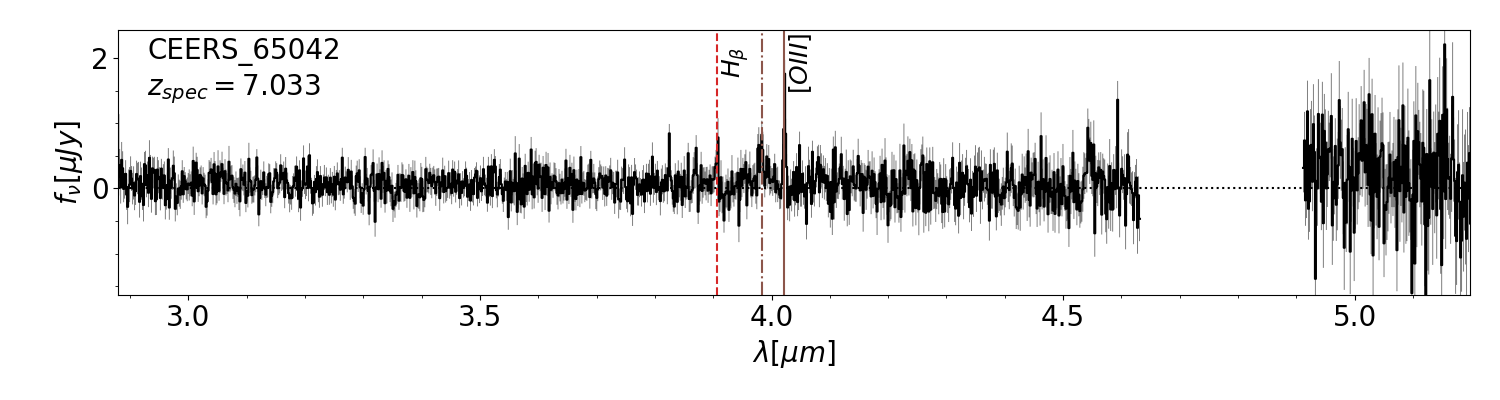}
    \includegraphics[width=\linewidth,trim={6 50 0 0},clip]{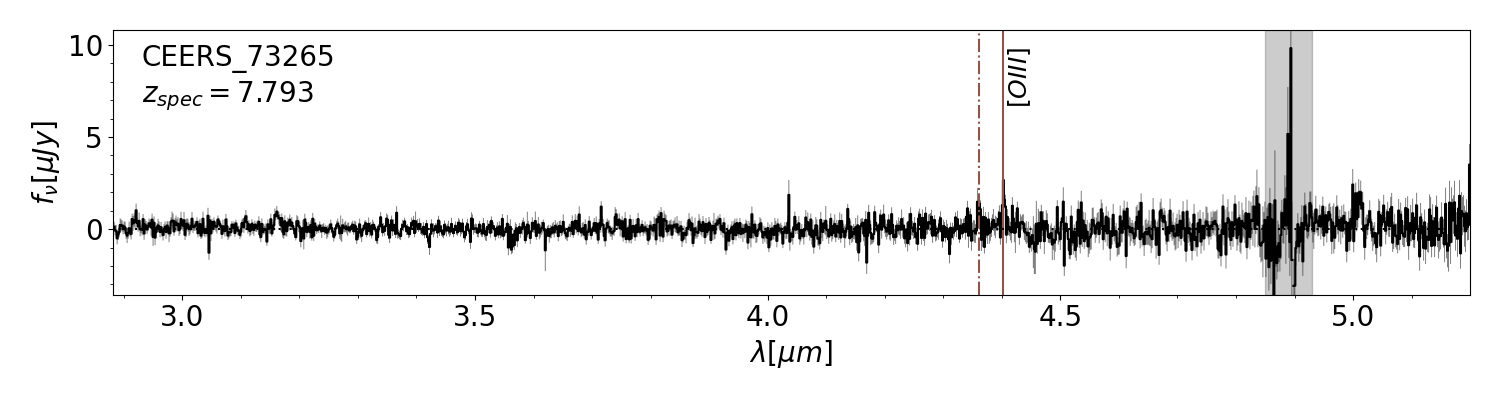}
    \includegraphics[width=\linewidth,trim={15 50 0 0},clip]{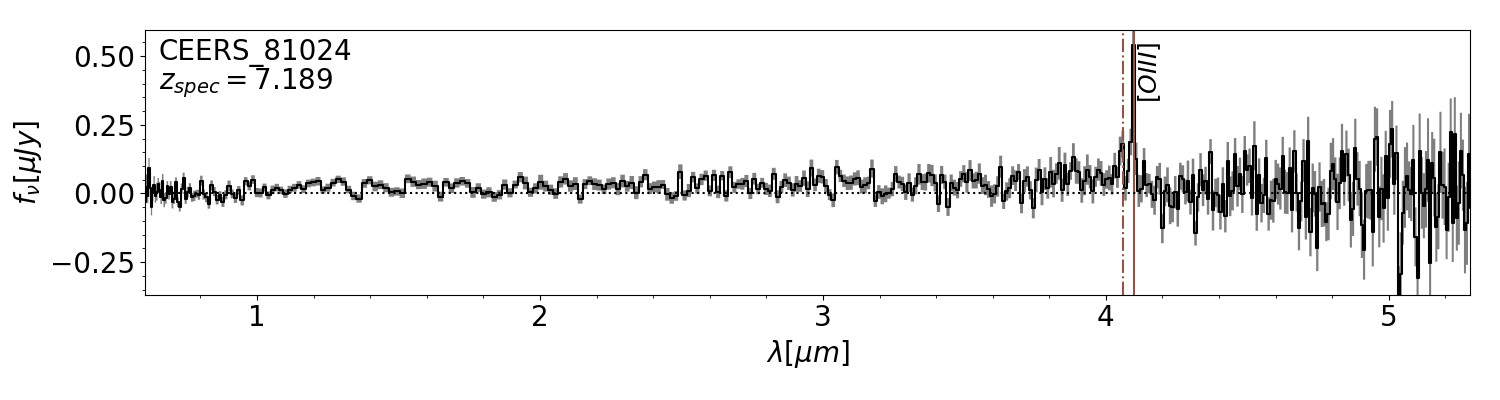}
    \includegraphics[width=\linewidth,trim={32 50 0 0},clip]{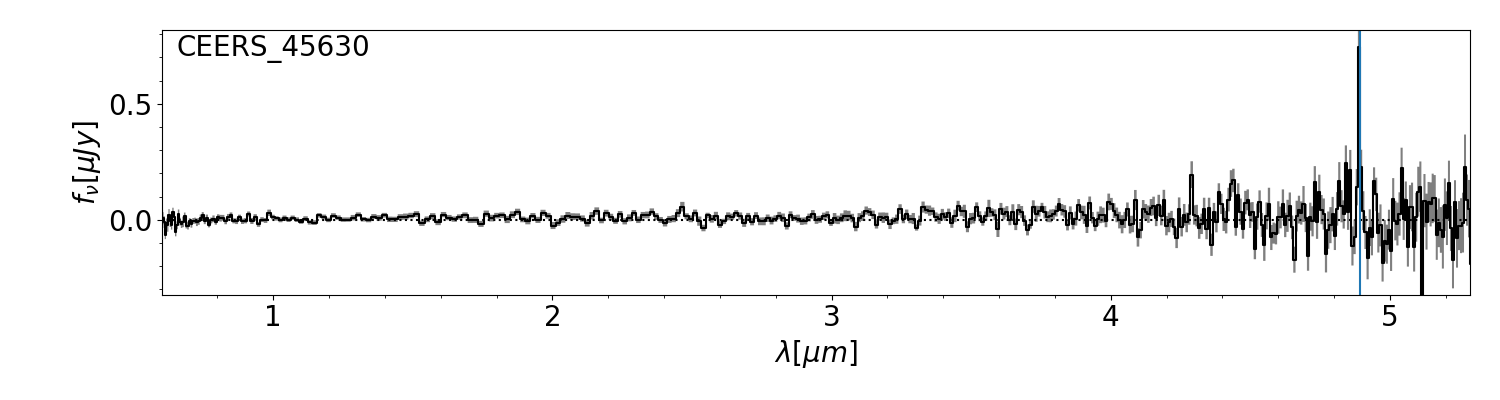}
    \includegraphics[width=\linewidth,trim={22 50 0 0},clip]{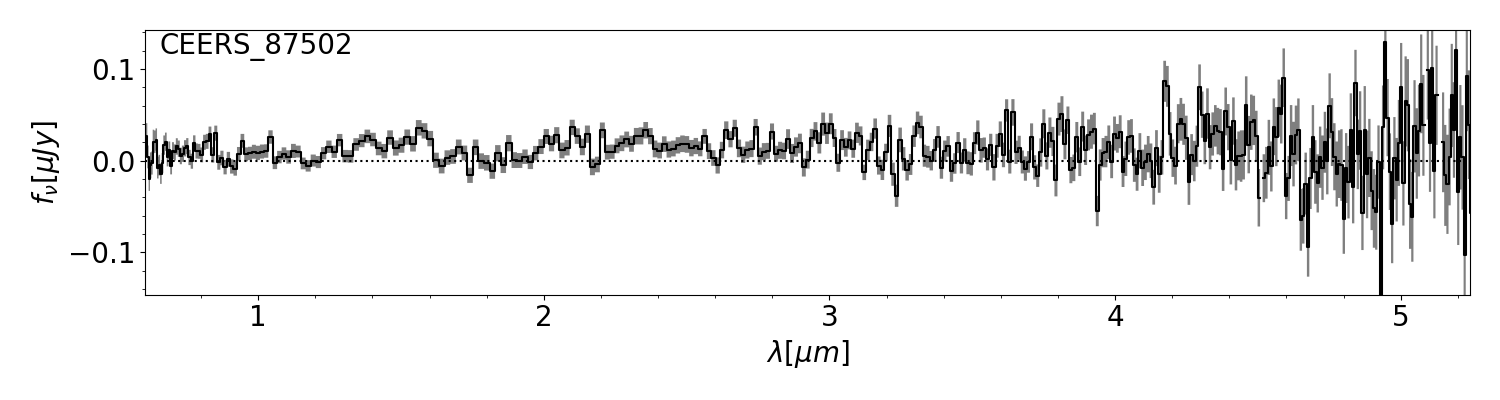}
    \includegraphics[width=\linewidth,trim={20 0 0 0},clip]{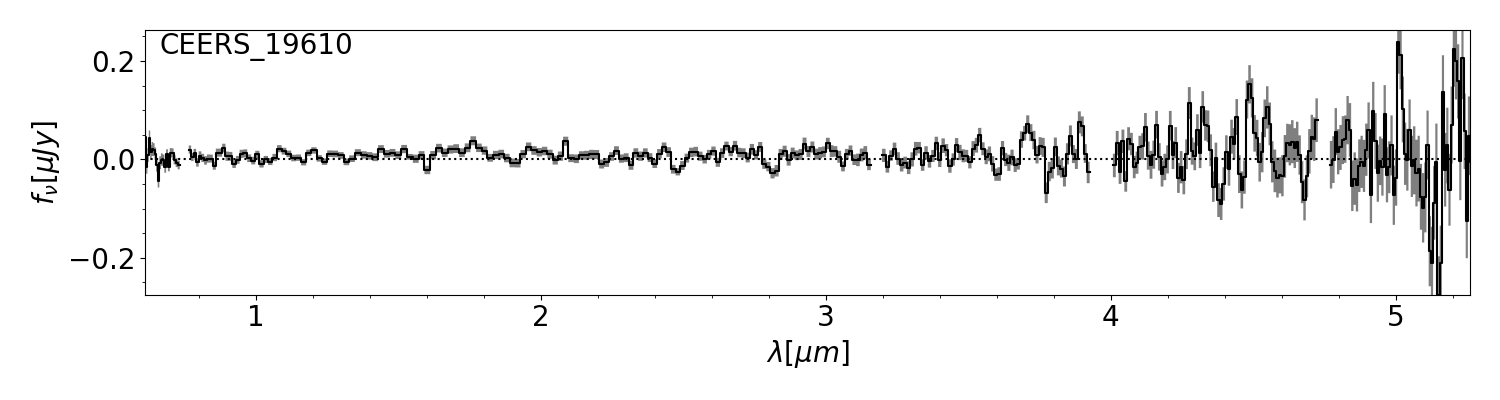}
    \caption{Sources in the HELM-$1\sigma$ sample with NIRSpec spectroscopic data. Vertical lines show nebular emission lines present in the spectra. In two objects (CEERS-45766 and CEERS-45630), only one emission line is visible preventing us to securely identify it. Other two spectra (CEERS-87502 and CEERS-19610) show no evident features. In the top left of each panel we report the source identification number and the derived spectroscopic redshift, if available.}
    \label{fig:s1_spectra}
\end{figure}
\subsection*{CEERS-45809}
This object was part of the HELM-$1\sigma$ sample, has visible $H_{\beta}$ and $[\ion{O}{III}]5007\AA$ lines (Fig. \ref{fig:s1_spectra}, first panel) and is inside the sample of extreme line emitters by \citet{Llerena2024}. The spectroscopic redshift is $z_{spec}=7.49$, but given that a single Hydrogen line is visible, it is not possible to derive the dust attenuation from the spectrum. We performed a spectro-photometric fit using \textsc{Bagpipes} and a setup similar to the one presented in Section \ref{sec:SEDfitting}, but considering a non-parametric SFH which is possible given that the spectroscopic redshift is available, as done by \citet{Bisigello2025a}. From the fit we derived a stellar mass of ${\rm log_{10}}(M_*/\rm M_{\odot})=7.94_{-0.09}^{+0.14}$ and a dust attenuation $A_V=0.9_{-0.1}^{+0.1}$~mag, so this galaxy is just outside the HELM definition defined in eq. \ref{eq:selection} and we removed it from the final sample.

\subsection*{CEERS-45766}
The NIRSpec PRISM spectrum of CEERS-45766, which is shown in Fig. \ref{fig:s1_spectra} (second row), shows a tentative detection of a single line, which is not sufficient to derive a robust spectroscopic redshift. In particular, the line is located at $\lambda=3.3$\mum with an observed flux of $(2.66\pm0.32)\times10^{-19}\,\rm erg/s/cm^2$, but there are no other robust lines. If the visible line is $H_{\alpha}$, the galaxy should be at $z=4.0$, while if the line is the $[\ion{O}{III}]5007\AA$ nebular line, the galaxy should be at $z=5.6$. In the first case, the noise is too high to derive any stringent lower limit in the dust attenuation from the non detection of $H_{\beta}$. In the second case, we would expect to observe $H_{\alpha}$ at $\lambda=4.33$\mum, but nothing is visible in that part of the spectrum, which is however quite noisy. 
Another possibility is that the line is Pa$_{\alpha}$ at a $z=0.76$, which would be consistent with the photometric redshift that ranges from $z=0.2$ to $z=1.4$, depending on the SED fitting setup. In this case we would expect to observe the $H_{\alpha}$ line at 1.15\mum, if the galaxy does not have an extreme dust attenuation, but no lines are visible in that area of the spectrum.

If we repeat the SED fitting described in Section \ref{sec:SEDfitting}, but fixing to one of the possible redshifts, we retrieved a dwarf galaxy ${\rm log_{10}}(M_*/\rm M_{\odot})\sim8.0$, but not a dusty one as $A_V=0.3\pm0.2$~mag for the $z\geq4$ solutions. If instead we fixed the the redshift to $z=0.76$ we retrieved a less massive galaxy ${\rm log_{10}}(M_*/\rm M_{\odot})\sim7.2$ with $A_V=1\pm0.5$~mag, which is however not enough to justify the completely absorb the $H_{\alpha}$ line. Considering these results, we decided to remove this source from the HELM-$1\sigma$ sample.

\subsubsection*{CEERS-29979}
This spectrum was obtained from the Cycle 1 GO-2565 program (P.I. K. Glazebrook). We derived a spectroscopic redshift of $z = 3.788 \pm 0.004$ from the [\ion{O}{III}]5007\AA\xspace emission line. While the $H_{\alpha}$ line is also visible in the spectrum (Fig. \ref{fig:s1_spectra}), the $H_{\beta}$ line is not detected. Given the spectrum's low S/N ratio, we proceeded by performing the SED fit at a fixed spectroscopic redshift, including only the photometric data. From this fit, we derived a stellar mass of ${\rm log_{10}}(M_*/\rm M_{\odot})=8.1_{-0.1}^{+0-1}$ and $A_V=0.5_{-0.1}^{+0.1}$, placing this galaxy outside the HELM selection. We therefore remove this source from the sample.

\subsubsection*{CEERS-30063}
CEERS-30063 has been observed as part of the RUBIES survey with the medium resolution PRISM/CLEAR and G395M/F290LP disperser-filter combinations, but the latter is to noisy to identify any feature. The spectroscopic redshift is $z=5.307\pm0.006$ derived from the [\ion{O}{III}]5007\AA\xspace line. The $H_{\alpha}$ is also visible, but the S/N is quite limited (Fig. \ref{fig:s1_spectra}) preventing us to derive the dust attenuation directly from the spectrum. We performed a spectro-photometric fit, including only the part of spectrum around fluxes with S/N$>3$. From the fit we derive a stellar mass of ${\rm log_{10}}(M_*/\rm M_{\odot})=8.4_{-0.1}^{+0.1}$ and $A_V=0.2_{-0.2}^{+0.2}$, which move this galaxy outside the HELM selection. 

\subsubsection*{CEERS-19613}
CEERS-19613 has been observed as part of the RUBIES survey with the PRISM/CLEAR and G395M/F290LP disperser-filter combinations. The spectroscopic redshift is $z=5.9828\pm0.0001$ derived from the [\ion{O}{III}]5007\AA\xspace line in the highest resolution spectrum, but many other nebular line are visible, among with both $H_{\alpha}$ and $H_{\beta}$. From their ratio in the G395M/F290LP spectrum, we derived that the dust attenuation is $A_V=3.4_{-0.9}^{+1.0}$ using the \citet{Calzetti2000} reddening law, but it remains large ($A_V=2.5_{-1.1}^{+0.8}$) even considering the maximum contribution from the stellar absorption below the $H_{\beta}$ line. 
We performed a spectro-photometric fit, fixing the redshift to the spectroscopic value, allowing the dust-extinction to varies between the $1\sigma$ uncertainties derived from the spectrum and considering a non parametric SFH. We obtained a stellar mass of $log_{10}(M_*/\rm M_{\odot})=9.6_{-0.1}^{+0.1}$, which would place this galaxy just outside the HELM selection. However the code struggle to fit the data hitting the upper boundary for the ionization parameter. We remove this source from the HELM sample.

\subsubsection*{CEERS-65042} 
This source has been observed as part of the RUBIES survey only with the G395M/F290L configuration. It has a spectroscopic redshift of $z=7.0332\pm0.0005$ derived from the [\ion{O}{III}]5007\AA\xspace. In the spectrum it is visible a faint $H_{\beta}$ line, while the $H_{\alpha}$ is outside the wavelength range covered. We perform a spectro-photometric fit with non-parametric SFH obtaining a stellar mass of $log_{10}(M_*/\rm M_{\odot})=8.9_{-0.2}^{+0.1}$ and a dust attenuation of $A_V=1.1_{-0.2}^{+0.1}$, which move this object slightly outside the HELM selection.

\subsubsection*{CEERS-73265} 
This source has been observed as part of the RUBIES survey with the PRISM/CLEAR and the G395M/F290L configurations. However, the first one is affected by contamination issues, while the spectra of the second is affected by artifacts only at long wavelengths. We obtain a redshift of $z=7.793\pm0.001$ from the [\ion{O}{III}]5007\AA\xspace, but only the [\ion{O}{III}]5007 doublet is clearly visible. We perform a spectro-photometric fit with non-parametric SFH obtaining a stellar mass of $log_{10}(M_*/\rm M_{\odot})=8.7_{-0.2}^{+0.1}$ and a dust attenuation of $A_V=0.5_{-0.2}^{+0.2}$, which move this object outside the HELM selection.

\subsubsection*{CEERS-81024} 
This source has been observed as part of the RUBIES survey only with the PRISM/CLEAR configuration. It has a spectroscopic redshift of $z=7.189\pm0.002$ derived from the [\ion{O}{III}]5007\AA\xspace and only the [\ion{O}{III}]5007 doublet is clearly visible. We perform a spectro-photometric fit with non-parametric SFH obtaining a stellar mass of $log_{10}(M_*/\rm M_{\odot})=8.8_{-0.2}^{+0.2}$ and a dust attenuation of $A_V=0.6_{-0.2}^{+0.1}$, which move this object outside the HELM selection.

\subsubsection*{CEERS-45630} 
This source has been observed as part of the RUBIES survey with the PRISM/CLEAR configuration. Only a single line is visible in the spectrum and if it corresponds to the [\ion{O}{III}]5007\AA\xspace line it would place the galaxy at $z=8.772\pm0.002$. We perform a spectro-photometric fit with non-parametric SFH obtaining a stellar mass of $log_{10}(M_*/\rm M_{\odot})=8.3_{-0.2}^{+0.3}$ and a dust attenuation of $A_V=0.6_{-0.2}^{+0.1}$, which move this object outside the HELM selection.
%\textcolor{red}{[I need to verify that this is the galaxy that was pointed as there is another one in the general catalog that is slightly closer]}
\subsubsection*{CEERS-87502, CEERS-19610}
The NIRSpec PRISM spectra of CEERS-87502 and CEERS-19610, which are shown in Fig. \ref{fig:s1_spectra} (third and fourth rows), show no evident nebular emission lines nor stellar continuum, so no conclusive statements can be made on the nature of these objects.

\end{appendix}

\end{document}